\documentclass[pra,aps]{revtex4}

\usepackage{amsmath}
\usepackage{ulem}
\usepackage{bm}
\usepackage{graphics}
\usepackage[dvips]{color}
\usepackage{dcolumn}

\begin{document}
\title{Spin dynamics of a trapped spin-1 Bose Gas above the Bose-Einstein transition temperature}
\author{Yuki Endo and Tetsuro Nikuni}
\affiliation{Department of Physics, Faculty of Science, Tokyo University of Science, 
1-3 Kagurazaka, Shinjuku-ku, Tokyo, Japan, 162-8601}
\date{\today}

\begin{abstract}
We study collective spin oscillations in a spin-1 Bose gas above the Bose-Einstein transition temperature. Starting from the Heisenberg equation of motion, we derive a kinetic equation describing the dynamics of a thermal gas with the spin-1 degree of freedom. Applying the moment method to the kinetic equation, we study spin-wave collective modes with dipole symmetry. The dipole modes in the spin-1 system are classified into the three types of modes. Frequencies and damping rates of the dipole modes are obtained as functions of the peak density. We find that the damping rate is characterized by three relaxation times associated with collisions.
\end{abstract}

\maketitle
\section{Introduction} 
Since the experimental realization of Bose-Einstein condensation (BEC) in dilute atomic gases~\cite{Anderson1995,Davis1995}, there have been a large number of theoretical and experimental studies on ultracold atomic gases. Among other things, BEC of atoms with internal degrees of freedom attracted much attention. In particular, the dynamics of a spin-1 Bose gas has been extensively studied, since the MIT group have succeeded in confining ${\rm ^{23}Na}$ Bose-Einstein condensate in an optical trap (so called spin-1 BEC)~\cite{Stenger1998}. Such spin-1 BEC exhibits many intriguing phenomena because these spin components can exchange each other through spin-spin interaction. Although the strength of the spin-spin interaction is typically an order of magnitude smaller than that of the spin-independent interaction, the spin-spin interaction have a significant effect on dynamical properties of spin-1 Bose gases~\cite{Stenger1998,Kurn1998,Miesner1999,Chang2004,Erhard2004,Higbie2005,Sadler2006}

Most of the studies in the dynamics of spin-1 Bose gases mainly concentrated on near zero-temperature dynamics in a pure condensate~\cite{Law1998,Ohmi1998,Kurn1999,Zhang2003,Zhang2005,Zhang2005_2,Kronjager2005,Michael2007}. On the other hand, there have also been a growing interest in finite-temperature properties of the spinor BEC, taking into account the effect of a thermal component~\cite{Szepfalusy2002,Szirmai2003,Schmaljohann2004,Erhard2004,Kronjager2005,Petit2006}. In this paper, we study the spin dynamics of spin-1 Bose gases above the Bose-Einstein transition temperature ($T_{\rm BEC}$).

In fact, spin-1/2 Bose gases are known to exhibit the collective spin dynamics even well above $T_{\rm BEC}$~\cite{Johnson1984,Bigelow1989}. In JILA experiments~\cite{McGuirk2002}, a trapped dilute Bose gas of ${\rm ^{87}Rb}$ atoms with two hyper-fine states were used to study the dynamics of a spin-1/2 Bose gas. Even though the JILA experiments were done in relatively high-temperature regime $T\sim 2T_{\rm BEC}$, where the quantum degeneracy has little effect on the thermodynamic properties, this spin-1/2 system exhibited collective spin dynamics due to the exchange effect, such as spin segregation~\cite{Lewandowski2002,Williams2002,Fuchs2002,Oktel2002_2}, and spin-wave oscillations~\cite{McGuirk2002,Nikuni2002}. There have been many theoretical studies discussing collective spin oscillations in the spin-1/2 system~\cite{Pitaevskii_book,Williams2002,Fuchs2002,Oktel2002_2,Nikuni2002,Nikuni2003,Fuchs2005,Ragan2005,Mullin2006}, which showed very good agreement with the experiments~\cite{Hall1998,Lewandowski2002,McGuirk2003}. 

We also expect the spin-1 thermal gas to exhibit collective spin oscillations due to the spin-spin interaction~\cite{Szepfalusy2002,Szirmai2003}.  The purpose of the present paper is to investigate of the collective spin oscillations in a trapped spin-1 Bose gas above the Bose-Einstein transition temperature. We extend the work on the spin-1/2 system in Ref.~\cite{Nikuni2002} to the spin-1 system. Ref.~\cite{Nikuni2002} discussed the collective spin oscillations with dipole and quadrupole symmetries  for the spin-1/2 Bose gas, applying the moment method to the kinetic equation~\cite{Odelin1999,Khawaja2000}. In this paper we discuss the dipole spin oscillations in a spin-1 Bose gas. While in the spin-1/2 system, the components couple each other through the exchange effect, in the spin-1 system the components couple through intrinsic spin-spin interaction. We thus expect the spin-1 Bose gas to exhibit a richer collective spin dynamics.

 In Sec.II, we first derive a kinetic equation for the spin-1 Bose gas. In Sec.III, starting from the kinetic equation, we derive moment equations describing spin oscillations with dipole symmetry~\cite{Griffin1997,Nikuni2002}. Solving the moment equations, we obtain explicit analytical expressions for frequency and damping rate of dipole modes. Asymptotic expressions for frequency and damping rate in collisionless limit and in the hydrodynamic limit are also obtained.

\section{DERIVATION OF THE SPIN-1 KINETIC EQUATION}
In this section, we give a derivation of the spin-1 kinetic equation, by which we will discuss the dipole oscillation above $T_{\rm BEC}$ in Sec.\ref{Section_dipole}. 

We consider a Bose gas of atoms with spin $F=1$ in an optical trap (spin-1 Bose gas). Each atom has three hyper spin states; $m_F=1,0,-1$, so the single-particle state is expressed by a spinor wave function.
\begin{eqnarray}
\uline{\phi} \left( \textbf{r},t \right) = \begin{pmatrix}
                      \phi_1 \left( \textbf{r},t \right) \\
                      \phi_0 \left( \textbf{r},t \right)\\
                       \phi_{-1} \left( \textbf{r},t \right) \\
                      \end{pmatrix}=\phi_1 \left( \textbf{r},t \right)\left| 1 \rangle\right.+\phi_0 \left( \textbf{r},t \right)\left| 0 \rangle\right.+\phi_{-1} \left( \textbf{r},t \right)\left| -1 \rangle\right..
\end{eqnarray}
Bold-faced variables are vectors in coordinate space and the single underline indicates a spin variable. The single atom Hamiltonian is given by
\begin{eqnarray}
\uuline{H_0} \left( \textbf{r} \right) &=& \left[ - \frac{\hbar^2}{2m}{\nabla}_{\textbf{r}}^2 + V \left( \textbf{r} \right) \right] \uuline{1} \nonumber \\
	& & +g\mu_B \sum_{\alpha}B^{\alpha}\left( \textbf{r} \right) \uuline{S}^{\alpha} + \sum_{\alpha\beta}B_q^{\alpha\beta}\left( \textbf{r} \right) \uuline{Q}^{\alpha\beta},\label{single_atom_Hamiltonian}
\end{eqnarray}
where the double underline indicates a $3\times 3$ matrix, and $\alpha,\beta$ indicate spin components ${\it x,y,z}$. The first term describes the center-of-mass motion of an atom in a harmonic trap $V \left( \textbf{r} \right)= \frac{m}{2} \left( \omega_x^2x^2 + \omega_y^2y^2 + \omega_z^2z^2 \right)$, and $m$ is the mass of a single atom. The second term describes the linear Zeeman energy,
where $B^{\alpha}$ is the magnetic field of spin-$\alpha$ component, and $\uuline{S}^{\alpha} \left( \alpha = x,y,z \right)$ denotes the spin-1 matrix of spin-$\alpha$ component  
\begin{eqnarray}
\uuline{S}^x=\frac{1}{\sqrt{\mathstrut 2}} \begin{pmatrix}
                       0 & 1 & 0 \\
                       1 & 0 & 1 \\
                       0 & 1 & 0 \\
                      \end{pmatrix} 
,\ \ \ \ \uuline{S}^y=\frac{1}{\sqrt{\mathstrut2}} \begin{pmatrix}
                       0 & -i & 0 \\
                       i & 0 & -i \\
                       0 & i & 0 \\
                     \end{pmatrix}
,\ \ \ \ \uuline{S}^z= \begin{pmatrix}
                       1 & 0 & 0 \\
                       0 & 0 & 0 \\
                       0 & 0 & -1 \\
     \end{pmatrix} ,\label{103}
\end{eqnarray}
and {\it g} is Lande's {\it g} factor and $\mu_B$ is Bohr magnetron. In the third term, we have introduced a generalized form of an external field $B_q^{\alpha\beta}$ that is coupled to quadrupole operator $\uuline Q^{\alpha\beta}$, which is defined by 
\begin{eqnarray}
&&\uuline{Q}^{\alpha \beta }= \left( 1-\frac{1}{2} \delta _{\alpha \beta }  \right) \left( \uuline{S}^\alpha \uuline{S}^\beta + \uuline{S}^\beta \uuline{S}^\alpha - \frac{2}{3} \uuline{S}^2 \delta _{\alpha \beta }\right) .
\end{eqnarray}
In the spin-1 system, quadratic Zeeman effect described by {\it zz}-component $B^{zz}$ plays an important role, as observed in dephasing and quenching phenomena~\cite{Kronjager2005, Sadler2006, Michael2007} in the presence of an external bias field. In Eq.(\ref{single_atom_Hamiltonian}), we have introduced more general form of the tensor field $B_q^{\alpha\beta}$, which can express the quadratic Zeeman effect in the presence of magnetic field in an arbitrary direction or some kind of two-photon coupling field. In Sec. III, we will only consider the effect of $B_q^{zz}$.

Next, we consider the effective interaction for two $F=1$ atoms. The two-body interaction is modeled by a $\delta$-function pseudo potential, which is described in the form
\begin{eqnarray}
U \left( \textbf{r}_1\textbf,\textbf{r}_2 \right) = \delta \left( \textbf{r}_1 - \textbf{r}_2 \right) \left( g_0\uuline{1}\otimes\uuline{1} + g_2\sum_\alpha {\uuline{S_1^\alpha}} \otimes {\uuline{S^\alpha_2}}  \right),
\end{eqnarray}
where the coupling constants $g_0$ and $g_2$ are given by
\begin{eqnarray}
g_0=\frac{4\pi\hbar^2}{3m}\left(a_0+2a_2\right),\ \ \ \ g_2=\frac{4\pi\hbar^2}{3m}\left(a_2-a_0\right),
\end{eqnarray}
with $a_0$ and $a_2$ being the s-wave scattering lengths for collision channels with total spins ${\cal{F}}=0$ and $2$, respectively.

In second quantized form, the many-body Hamiltonian of this system is given by
\begin{eqnarray}
\hat{H}\left(\textbf{r}\right) &=& \sum_{ij} \int d\textbf{r} \hat \Psi _i^\dagger \left( \textbf{r} \right) \langle i \mid \uuline{H_0} \left( \textbf{r} \right) \mid j \rangle \hat \Psi _j  \left( \textbf{r} \right) \notag \\
	& & +\frac{g_0}{2}\sum_{ij}\int d \textbf{r} \hat \Psi _i^\dagger  \left( \textbf{r} \right) \hat \Psi _j^\dagger  \left( \textbf{r} \right) \hat \Psi _j  \left( \textbf{r} \right) \hat \Psi _i  \left( \textbf{r}\right) \notag \\
	& & +\frac{g_2}{2} \sum_{iji^\prime j^\prime }\sum_{\alpha } \int d \textbf{r} \hat \Psi _i^\dagger  \left( \textbf{r} \right) \hat \Psi _{i^\prime }^\dagger  \left( \textbf{r}\right) S_{ij}^\alpha S_{i^\prime j^\prime } ^\alpha \hat \Psi _{j^\prime } \left( \textbf{r}\right) \hat \Psi _j \left( \textbf{r}\right),    
\end{eqnarray}
where $\hat \Psi_i^\dagger \left( \textbf{r} \right)$ is the Bose field operator satisfying the commutation relation
\begin{eqnarray}
\left[ \hat \Psi_i \left( \textbf{r} \right), \hat\Psi_j^\dagger \left( \textbf{r}^\prime \right) \right] = \delta_{ij} \delta \left( \textbf{r} -\textbf{r}^\prime \right),
\end{eqnarray}
and $i, j$ indicate the hyperfine spin states $m_F=1,0,-1$ and the hat $\left(\ \hat{}\ \right)$ indicates a second quantized operator. 

We next define the time evolution of the system in terms of the statistical density operator, and introduce the Heisenberg representation for the field operators. The dynamics of the system is described by the statistical density operator $\hat\rho\left(t\right)$, from which one can obtain the expectation value of an arbitrary operator $\hat O$ (which has no explicit time dependence)
\begin{eqnarray}
\langle \hat O\rangle_t={\rm tr}\ \hat\rho\left( t\right)\hat O.\label{100}
\end{eqnarray}
The state of the many-body system evolves in time according to
\begin{eqnarray}
i\hbar\frac{d\hat\rho\left(t\right)}{dt}=\left[\hat H,\hat\rho\left(t\right)\right].\label{101}
\end{eqnarray}
With Eqs.(\ref{100}) and (\ref{101}), the equation of motion for the quantity $\langle\hat O\rangle_t$ is given by
\begin{eqnarray}
i\hbar\frac{d\langle \hat O\rangle_t}{dt}=\langle\left[\hat O,\hat H\right]\rangle_t.
\end{eqnarray}

It is convenient to introduce the time evolution operator $\hat {\cal U}\left(t,t_0\right)$, which obeys the equation of motion
\begin{eqnarray}
i\hbar\frac{d\hat{\cal U}\left(t,t_0\right)}{dt}=\hat H\hat{\cal U}\left(t,t_0\right),
\end{eqnarray}
with $\hat{\cal U}\left(t_0,t_0\right)=1$. Here $t_0$ is the time at which the initial nonequilibrium density matrix $\hat\rho\left(t_0\right)$ is specified. One can then express the time evolution of $\hat\rho\left(t\right)$ as
\begin{eqnarray}
\hat\rho\left(t\right)=\hat{\cal U}\left(t,t_0\right)\hat\rho\left(t_0\right)\hat{\cal U}^\dagger\left(t,t_0\right).
\end{eqnarray}
Thus, the time evolution of $\langle\hat O\rangle_t$ can be written as
\begin{eqnarray}
\langle\hat O\rangle_t={\rm tr}\ \hat{\cal U}\left(t,t_0\right)\hat\rho\left(t_0\right)\hat{\cal U}^\dagger\left(t,t_0\right)\hat O ={\rm tr}\ \hat\rho\left(t_0\right)\hat{\cal U}^\dagger\left(t,t_0\right)\hat O \ \hat{\cal U}\left(t,t_0\right)\equiv\langle\hat O\left(t\right)\rangle,
\end{eqnarray}
where $\hat O\left(t\right)\equiv\hat{\cal U}^\dagger\left(t,t_0\right)\hat O\ \hat{\cal U}\left(t,t_0\right)$ is the operator in the Heisenberg picture, which obeys the Heisenberg equation of motion
\begin{eqnarray}
i\hbar\frac{\partial\hat O\left(t\right)}{\partial t}=\left[\hat O\left(t\right),\hat H\left(t\right)\right].
\end{eqnarray}
The Heisenberg equation of motion for the Bose field operator is given by
\begin{eqnarray}
i\hbar \frac{\partial}{\partial t} \hat \Psi_i \left( \textbf{r},t \right) = \left[ \hat\Psi_i \left( \textbf{r},t \right),\hat H\left(t \right) \right].
\end{eqnarray}

\subsection{The general kinetic equation of the nonequilibrium system}
In order to describe time evolution of the nonequilibrium system, we introduce the Wigner operator
\begin{eqnarray}
\hat {W}_{ij} \left( \textbf{p},\textbf{r} \right) = \int d \textbf{r}^\prime e^{i\textbf{p} \cdot \textbf{r}^\prime / \hbar }%
          \hat \Psi ^\dagger _{j}\left( \textbf{r} + \textbf{r}^\prime /2 \right)\hat \Psi _{i}\left( \textbf{r} - \textbf{r}^\prime /2 \right),
\end{eqnarray}
and define the semi-classical distribution function
\begin{eqnarray}
W_{ij}\left( \textbf{p},\textbf{r},t \right) \equiv \langle i\mid \uuline{W}\left( \textbf{p},\textbf{r} \right)\mid j \rangle\equiv {\rm tr}\hat \rho \left( t \right) \hat W_{ij}\left( \textbf{p},\textbf{r} \right),
\end{eqnarray}
where $\hat \rho \left( t_0 \right)$ is the statistical density operator. Knowledge of this function allows one to calculate various nonequilibrium physical quantities, such as the local density given by
\begin{eqnarray}
\uuline{n} \left( \textbf{r},t \right) = \int \frac{d \textbf{p}}{\left( 2\pi \hbar \right)^3} \uuline{W} \left( \textbf{p},\textbf{r},t \right),
\end{eqnarray}
where
\begin{eqnarray}
\langle i \mid \uuline{n} \left( \textbf{r},t \right) \mid j \rangle = n_{ij} \left( \textbf{r},t \right) \equiv \langle \hat \Psi_i^\dagger \left( \textbf{r},t \right) \hat \Psi_j \left( \textbf{r},t \right) \rangle
\label{1}.
\end{eqnarray}
From this local density, the total density, magnetization and quadrupole moment are given by
\begin{eqnarray}
&& n \left( \textbf{r},t \right) \equiv {\rm Tr}\ \uuline{n} \left( \textbf{r},t \right), \\
&& M^\alpha \left( \textbf{r},t \right) \equiv {\rm Tr}\ \uuline{S}^\alpha\uuline{n} \left( \textbf{r},t \right), \\
&& A^{\alpha\beta} \left( \textbf{r},t \right) \equiv {\rm Tr}\ \uuline{Q}^{\alpha\beta}\uuline{n} \left( \textbf{r},t \right) ,
\end{eqnarray}
where  ``${\rm Tr}$'' denotes the spin trace.

In order to derive a kinetic equation for $W_{ij}\left(\textbf{p},\textbf{r},t\right)$, it is useful to separate the Hamiltonian into a part that describes the mean-field dynamics $\hat H_{\rm MF}$ and the remaining part $\hat H^\prime$
\begin{eqnarray}
\hat H \left( t \right) = \hat H_{\rm MF} \left( t \right) + \hat H^\prime \left( t \right) ,\label{5}
\end{eqnarray}
where the leading mean-field Hamiltonian is 
\begin{eqnarray}
\hat H_{\rm MF} \left( t \right) = \sum_{ij} \int d \textbf{r} \hat \Psi ^\dagger _{i} \left(\textbf{r}\right)\langle i \mid \uuline{H_{0}} \left( \textbf{r},t \right) +\uuline{U_{1}} \left( \textbf{r},t \right) \mid j \rangle\hat \Psi _{j} \left( \textbf{r} \right),
\end{eqnarray}
 with $\uuline{U_1}\left(\textbf{r},t\right)$ being the Hartree-Fock (HF) mean-field potential
 \begin{eqnarray}
 \uuline{U_{1}} \left( \textbf{r},t \right) = g_{0}\uuline{n}\left( \textbf{r},t \right) + g_{0}n\left( \textbf{r},t \right) \uuline{1} + g_{2} \sum_{\alpha }\uuline{S}^\alpha \uuline{n}\left( \textbf{r},t \right)\uuline{S}^\alpha + g_{2} \sum_{\alpha } M^\alpha \left( \textbf{r},t \right) \uuline{S}^\alpha \label{3}.
\end{eqnarray}
The perturbation contribution in $H^\prime \left(t\right)$ is 
\begin{eqnarray}
\hat H^\prime \left( t \right) &=& \frac{1}{2} \sum_{iji^\prime j^\prime}\int d\textbf{r} \hat \Psi ^\dagger _{i} \left( \textbf{r} \right) \hat \Psi ^\dagger _{i^\prime} \left( \textbf{r} \right)V_{ii^\prime jj^\prime}\hat \Psi _{j^\prime} \left( \textbf{r} \right) \hat \Psi _{j} \left( \textbf{r} \right)\nonumber \\
& & -
\sum_{ij} \int d \textbf{r} \hat \Psi ^\dagger _{i} \left(\textbf{r}\right)\langle i \mid\uuline{U_{1}} \left( \textbf{r},t \right) \mid j \rangle\hat \Psi _{j} \left( \textbf{r} \right), 
\end{eqnarray}
where for convenience, we have defined the following matrix describing the two-body interaction
\begin{eqnarray}
V_{ii^\prime jj^\prime} \equiv g_0 \delta_{ij^\prime}\delta_{i^\prime j} + g_2 \vec S_{ij} \cdot \vec S_{i^\prime j^\prime} .
\end{eqnarray}
In writing Eq.(\ref{5}), $H^\prime\left(t\right)$ is viewed as a perturbation to the zeroth-order Hamiltonian $H_{\rm MF}\left(t\right)$. Noting that $\uuline{U_1}\left(t\right)$ is the self-comsistent HF mean-field, $H_{\rm MF}\left(t\right)$ defines excitations of the system at the level of the time-dependent HF approximation. It will be shown later $H_{\rm MF}\left(t\right)$ contributes to the free-streeming term of the kinetic equation, while $H^\prime\left(t\right)$ accounts for binary collisions between atoms.

With the above definitions, the equation of motion for $W_{ij}\left(\textbf{p},\textbf{r},t\right)$ is written as
\begin{eqnarray}
\frac{\partial }{\partial t} W_{ij} \left( \textbf{p},\textbf{r},t \right) &=& \frac{1}{i\hbar }{\rm tr}\hat \rho \left( t \right)\left[ \hat W_{ij}\left( \textbf{p},\textbf{r} \right) ,\hat H \left( t \right) \right] \nonumber \\
        &=&\frac{1}{i\hbar }{\rm tr}\hat \rho \left( t \right) \left[ \hat W_{ij}\left( \textbf{p},\textbf{r}\right) ,\hat H_{\rm MF} \left( t \right) \right] +\frac{1}{i\hbar }{\rm tr}\hat \rho \left( t \right) \left[ \hat W_{ij}\left( \textbf{p},\textbf{r} \right) ,\hat H^\prime \left( t \right) \right] \label{2}.
\end{eqnarray}
For calculating the first term of Eq.(\ref{2}), we assume that the macroscopic variables are slowly varying in space, and make use of the approximation
\begin{eqnarray}
n_{ij} \left( \textbf{r} - \textbf{r}^\prime, t \right)  \approx n_{ij} \left( \textbf{r},t \right) - \textbf{r}^\prime \cdot \nabla_\textbf{r} n_{ij} \left( \textbf{r},t \right).
\end{eqnarray}
We also use analogous approximations for $V$, $B^\alpha$, and $B_q^{\alpha\beta}$.

With these approximations, we obtain the kinetic equation for the distribution function
\begin{eqnarray}
\frac{\partial \uuline{W}}{\partial t} + \frac{\textbf{p}}{m}\cdot \nabla _{r} \uuline{W} - \frac{1}{2} \left\{ \nabla _{r} \uuline{U},\nabla _{p} \uuline{W} \right\}- \frac{i}{\hbar } \left[ \uuline{W},\uuline{U}\right] = \uuline{I} ,\label{4}
\end{eqnarray}
where $\left[ \ \ , \ \ \right]$ and $\left\{ \ \ , \ \ \right\}$ represent the commutator and anticommutator for the $3\times 3$ matrices. The left hand side of this equation represents the free-streaming term which is derived from the first term on the right hand side of Eq.(\ref{2}), while the right hand side of this equation represents the collision term which is derived from the last term of Eq.(\ref{2}), respectively.
We have defined the effective potential using Eq.(\ref{3}) as
\begin{eqnarray}
\uuline{U} \left( \textbf{r},t \right)  \equiv \uuline{U_0} \left( \textbf{r},t \right) +\uuline{U_1} \left( \textbf{r},t \right),\label{Unoteigi}
\end{eqnarray}
where
\begin{eqnarray}
\uuline{U_{0}} \left( \textbf{r},t \right) = V \left( \textbf{r},t \right) \uuline{1} +g\mu _{B} \sum_{\alpha }B^\alpha  \left( \textbf{r},t \right)%
      \uuline{S}^\alpha + \sum_{\alpha \beta }B_{q}^{\alpha \beta } \left( \textbf{r},t \right) \uuline{Q}^{\alpha \beta }.\label{U_0noteigi}
\end{eqnarray}
The term on the right-hand side of Eq.(\ref{4}) is the collision term, which is given by 
\begin{eqnarray}
\langle i \mid \uuline{I} \mid j \rangle = \frac{1}{i\hbar }{\rm Tr} \hat \rho \left( t_0 \right) \left[ \hat W_{ij}\left( \textbf{p},\textbf{r},t \right) ,\hat H^\prime \left( t \right) \right] .\label{7}
\end{eqnarray}
The reduction of this term to the form of a binary collision integral is a lengthy exercise. In Appendix \ref{Appendix_Collision}, we provide a detailed derivation of the collision integral using perturbation technique. The final result is written as 
\begin{eqnarray}
\uuline{I} &=& \left.\frac{\partial \uuline{W}}{\partial t}\right|_{coll} + \frac{i}{\hbar }\left[ \uuline{W}\left(\textbf{p}\right),\delta\uuline{U_n} \left( \textbf{p} \right) \right] ,
\end{eqnarray}
where $\frac{\partial \uuline{W}}{\partial t}\mid_{coll}$ is the collision integral and $\delta \uuline{U_n}$ is the second-order effective potential. Explicit expressions of these terms are given in Eqs.(\ref{coll2}) and (\ref{coll}) in Appendix \ref{Appendix_Collision}. For a dilute Bose gas considered in this paper, one can neglect $\delta \uuline{U_n}$~\cite{Zaremba1999,Nikuni2003}. 

In summary, we have derived the kinetic equation for the distribution function $\uuline{W}$ for the spin-1 Bose gas: 
\begin{eqnarray}
\frac{\partial \uuline{W}}{\partial t} + \frac{\textbf{p}}{m}\cdot \nabla _{r} \uuline{W} - \frac{1}{2} \left\{ \nabla _{r} \uuline{U},\nabla _{p} \uuline{W} \right\}- \frac{i}{\hbar } \left[ \uuline{W},\uuline{U}\right] = \left.\frac{\partial \uuline{W}}{\partial t}\right|_{coll},\label{9}
\end{eqnarray}
where the total effective potential $\uuline{U}$ is defined in Eqs.(\ref{3}), (\ref{Unoteigi}), and (\ref{U_0noteigi}), and explicit expression for the collision integral is given in Eq.(\ref{1001}). The above kinetic equation is one of the main results in this paper. Starting from this kinetic equation, we will discuss the collective modes of the spin-1 system in Section III.

\section{MOMENT METHOD FOR A TRAPPED SPIN-1 GAS}\label{Section_dipole}
In this section, we consider small amplitude spin oscillations around the fully polarized state where all the spins point up, which implies that all atoms are initially in the state $\mid F=1,m_F =1 \rangle$. We linearize the kinetic equation Eq.(\ref{9}) around this initial state. The initial distribution function $\uuline{W}^0 \left( \textbf{p},\textbf{r} \right)$ is given by 
\begin{eqnarray}
\left\{
\begin{array}{l}
 W^0_{11} \left( \textbf{p},\textbf{r} \right) = f_0 \left( \textbf{p},\textbf{r} \right) ,\\
 W^0_{ij} \left( \textbf{p},\textbf{r} \right) = 0\ \ \ \ {\rm otherwise} .
\end{array}
\right.\label{equilibrium_distribution}
\end{eqnarray}
Here we assume that the temperature is approximately twice that needed for Bose-Einstein condensation $T_{\rm BEC}$, and thus the initial distribution $f_0\left(\textbf{p},\textbf{r}\right)$ in Eq.(\ref{equilibrium_distribution}) is given by the Maxwell-Boltzmann distribution. Moreover, we assume that ${\it g_in/k}_{\rm B}T\ll 1$ and thus the mean-field does not affect the center-of-mass motion. With these assumptions, the initial distribution in Eq.(\ref{equilibrium_distribution}) is given by
\begin{eqnarray}
f_0 \left( \textbf{p},\textbf{r} \right) = \exp \left\{ -\beta \left[\frac{p^2}{2m} + V \left( \textbf{r} \right) - \mu _0 \right]\right\} .\label{for_Appendix B}
\end{eqnarray} 
The initial density is given by
\begin{eqnarray}
n_0\left( \textbf{r} \right)&=& \int\frac{d\textbf{p}}{\left(2\pi\hbar\right)^3}f_0 \left( \textbf{p},\textbf{r} \right) =\frac{1}{\lambda_{\rm th}^3}\exp \left\{ -\beta \left[ \frac{p^2}{2m} + V \left( \textbf{r} \right) - \mu _0 \right] \right\} \nonumber \\
&=& n_0 \left( 0 \right) e^{-\beta V \left( \textbf{r} \right)},
\end{eqnarray}
where $\lambda_{\rm th} = \left( 2\pi\hbar^2/mk_{\rm B}T\right)^{1/2}$ is the thermal de Broglie wave length, $\beta=1/k_{\rm B}T$ is the inverse temperature, $\mu_0$ is initial chemical potential, and $n_0\left( 0 \right)=e^{\beta \mu_0}/\lambda_{\rm th}^3$ is the initial density at the center of the trap potential. We then substitute $\uuline{W}\left( \textbf{p},\textbf{r},t \right) = \uuline{W}^0\left( \textbf{p},\textbf{r}\right)+\delta\uuline{W}\left( \textbf{p},\textbf{r},t \right)$ and $\uuline{U}\left( \textbf{p},\textbf{r},t \right) = \uuline{U}^0\left( \textbf{p},\textbf{r}\right)+\delta\uuline{U}\left( \textbf{p},\textbf{r},t \right)$ into Eq.(\ref{9}) to obtain the linearized kinetic equation. Since we assume $T\gg T_{\rm BEC}$, we neglect the Bose enhancement factor taking $W_{ij}^> \rightarrow \delta_{ij}$ in the collision integral Eq.(\ref{1001}). 

For simplify, we assume that the uniform magnetic field is applied in the $z$ direction. Then, the initial effective potential is given by
\begin{eqnarray}
\uuline{U}^0 \left( \textbf{r} \right) &=&V \left( \textbf{r} \right) \uuline{1} +  \epsilon _0 \uuline{S}^z  + \epsilon _1\uuline{Q}^{zz} \nonumber  \\
                       &+& W_0 \uuline{n_0} \left( \textbf{r} \right) + W_0{n}_0 \left( \textbf{r} \right)  \uuline{1} + W_2 \sum_{\alpha} \uuline{S}^{\alpha} \uuline{n_0} \left( \textbf{r} \right)  \uuline{S}^{\alpha} + W_2 \sum_{\alpha} M^{\alpha}_0  \left( \textbf{r} \right) \uuline{S}^{\alpha},
\end{eqnarray}
where the linear Zeeman energy is $\epsilon_0 = g\mu_{\rm B}B$. The quadratic Zeeman energy $\epsilon_1 = \frac{g^2\mu_{\rm B}^2}{16h\nu_{\rm hf}}B^2$ corresponds to $B_q^{zz}$~\cite{Stenger1998}, where $\nu_{\rm hf}$ is the hyperfine splitting. As noted above, we consider the high temperature region $g_in_0\left(0\right)/k_{\rm B}T\ll1$, where the gradient of the effective potential can be approximated as
\begin{eqnarray}
\nabla _{\textbf{r}} \uuline{U}^0 \left( \textbf{r} \right) &\approx& \nabla_{\textbf{r}} V \left( \textbf{r} \right) \uuline{1} + \nabla_{\textbf{r}} \epsilon_0 \uuline{S}^z + \nabla_{\textbf{r}} \epsilon_1 \uuline{Q}^{zz} \nonumber \\
	&=& \nabla_{\textbf{r}}V\left(\textbf{r}\right)\uuline{1}.
\end{eqnarray} 
Since the external potential and magnetic field are assumed to be time-independent, the fluctuation of the effective potential is given by
\begin{eqnarray}
\delta\uuline{U}\left(\textbf{r},t\right) = g_0\delta\uuline{n}\left(\textbf{r},t\right)+g_0\delta n\left(\textbf{r},t\right)\uuline{1}+g_2\sum_{\alpha}\uuline{S}^\alpha\delta \uuline{n}\left(\textbf{r},t\right)\uuline{S}^\alpha +g_2\sum_{\alpha}\delta M^\alpha\left(\textbf{r},t\right)\uuline{S}^\alpha,
\end{eqnarray}
where we have defined the fluctuation of magnetization as $\delta M^\alpha \left( \textbf{r},t \right) \equiv {\rm Tr} \uuline{S}^\alpha\delta\uuline{n} \left( \textbf{r},t \right)$. Considering the high-temperature regime $g_in_0\left(0\right)/k_{\rm B}T\ll1$, and assuming small amplitude fluctuations $\delta n\ll n_0$, we approximate $\nabla_\textbf{r}\delta \uuline{U}\approx 0$.

With these approximations, we arrive at the linearized kinetic equation for $\delta\uuline{W}$,
\begin{eqnarray}
\frac{\partial}{\partial t}\uuline{\delta W} &+& \frac{\textbf{p}}{m}\cdot\nabla_{\textbf{r}}\uuline{\delta W}
 - \nabla_{\textbf{r}}V \left( \textbf{r} \right)\cdot\nabla_{\textbf{p}}\uuline{\delta W}
\nonumber \\
&-&\frac{i}{\hbar}\epsilon_0\uuline{\delta W}^A
 -\frac{i}{\hbar}\epsilon_1\uuline{\delta W}^B \nonumber \\
&-&\frac{i}{\hbar}g_0 \left[ n_0 \left( \textbf{r} \right)\uuline{\delta W}^C+f_0\left(\textbf{p},\textbf{r}\right)\uuline{\delta n}^C\right]
-\frac{i}{\hbar}g_2 \left[ n_0 \left( \textbf{r} \right)\uuline{\delta W}^D+f_0\left(\textbf{p},\textbf{r}\right)\uuline{\delta n}^D\right]
 \nonumber \\                      
&=&\left.\frac{\partial \uuline{\delta W }}{\partial t}\right|_{coll},     \label{10}
\end{eqnarray}
where we have introduced simplified notations for several matrices, which are defined in Appendix \ref{Appendix_Moment}. The linearized collision integral $\left.\partial\delta W /\partial t\right|_{coll}$ is given in Appendix \ref{Appendix_Relaxation}.

We take moment of the linearized kinetic equation Eq.(\ref{10}) for general quantity $\chi_a \left( \textbf{p},\textbf{r} \right)$~\cite{Odelin1999,Khawaja2000,Nikuni2002}. We define the expectation value of the fluctuation from the initial state and the part of this which depend on  the collision term for $\chi _a\left( \textbf{p},\textbf{r} \right)$ as
\begin{eqnarray}
&& \langle \chi_a \rangle _{ij} = \frac{1}{N} \int d \textbf{r} \int \frac{ d \textbf{p}}{\left( 2 \pi \hbar \right) ^3 } \chi_a  \delta W_{ij}\left(\textbf{p},\textbf{r},t\right) \label{chi},\label{X1}\\
&&  \langle \chi_a \rangle ^{eq} _{ij} = \frac{1}{N} \int d \textbf{r} \int \frac{ d \textbf{p}}{\left( 2 \pi \hbar \right) ^3 } \chi_a f_0\left(\textbf{p},\textbf{r}\right) ,\label{X2}\\
&&  \langle \chi_a \rangle ^{coll} _{ij} = \frac{1}{N} \int d \textbf{r} \int \frac{ d \textbf{p}}{\left( 2 \pi \hbar \right) ^3 } \chi_a \left.\frac{\partial \delta W_{ij}}{\partial t} \right|_{coll} \label{X3}.
\end{eqnarray}
Using the above moments, we can derive the general moment equation for $\langle \chi_a\left(\textbf{p},\textbf{r}\right)\rangle$ from the linearized kinetic equation Eq.(\ref{10}).

In general, the collision terms couple several components of the matrix $ \uuline{\langle{\chi_a }\rangle}$, and thus the density {\it n}, magnetization $M^\alpha$ and quadrupole moment $A^{\alpha\beta}$ are coupled to each other. We thus obtain form moment equations in matrix. With the appropriate chose of the function $\chi_a \left( \textbf{p},\textbf{r}\right)$, one can discuss several collective modes for the spin-1 Bose gas. In this paper, we consider oscillations of the following three quantities: $\chi_0=1$, $\chi_1=x$, $\chi_2=p_x/m$.

\subsection{Precession mode}
First, we consider the precession mode, taking $\chi_0=1$. Inserting $\chi_0$ into the general moment equation, we obtain the moment equation:
\begin{eqnarray}
\frac{d}{dt}\uuline{\langle\chi_0\rangle}-\frac{i}{\hbar}\epsilon_0\uuline{\langle \chi_0\rangle}^A-\frac{i}{\hbar}\epsilon_1\uuline{\langle\chi_0\rangle}^{B}-i\omega_{\rm MF}^{\left(2\right)}\uuline{\langle\chi_0\rangle}^{D}=\uuline{\langle\chi_0\rangle}^{coll},\label{A1}
\end{eqnarray}
where we have introduced simplified notations for several matrices, which are defined in Appendix \ref{Appendix_Moment}. The mean-field frequency is defined as
\begin{eqnarray}
\hbar \omega_{\rm MF}^{\left( {\cal F} \right)} \equiv \frac{g_{{\cal F}}n_0\left(0 \right)}{2\sqrt 2},\label{12}
\end{eqnarray}
where ${\cal F}$ denotes the total angular moment 0 or 2. To obtain a closed set of equations, we must truncate the collision terms $\uuline{\langle\chi_0\rangle}^{coll}$ by expanding the fluctuations in the distribution function $\delta W_{ij}$ in powers of position and momentum. For precession modes and dipole modes (which we will consider in the next subsection), it is sufficient to take the following form: 
\begin{eqnarray}
\delta W_{ij} \left( \textbf{p},\textbf{r},t \right) = f_0 \left(\textbf{p},\textbf{r} \right) \left[ \alpha_0^{ij} \left( t \right) + \alpha_1^{ij} \left(t \right)x+\alpha_2^{ij}\left(t\right)p_x \right]. \label{13}
\end{eqnarray}
The coefficients in the expansion can be related back to the set of moments using Eq.(\ref{chi}); $\alpha_0^{ij} =\langle\chi_0\rangle_{ij}$, $\alpha_1^{ij}=\beta m \omega_x^2\langle\chi_1\rangle_{ij}$, $ \alpha_2^{ij}=\beta\langle\chi_2\rangle_{ij}$. We linearize Eq.(\ref{1001}) in $\delta \uuline{W}\left(\textbf{p},\textbf{r},t\right)$ and obtain the collision term as
\begin{eqnarray}
\uuline{\langle\chi_0\rangle}^{coll}=2\gamma_0\begin{pmatrix}
	-2\langle\chi_0\rangle_{-1-1} & \ 3\langle\chi_0\rangle_{0-1} & -6\langle\chi_0\rangle_{1-1} \\
	\ 3\langle\chi_0\rangle_{-10} & \ 4\langle\chi_0\rangle_{-1-1} & -3\langle\chi_0\rangle_{0-1} \\
	-6\langle\chi_0\rangle_{-11} & -3\langle\chi_0\rangle_{-10} & -2\langle\chi_0\rangle_{-1-1}
	\end{pmatrix}.\label{A2}
\end{eqnarray}
Some details on the derivation of the above expression and the definition of the relaxation rate $\gamma_0$ are given in Appendix \ref{Appendix_Relaxation}. 

Using the expression for the chemical potential $e^{\beta\mu_0}=N\left(2\pi\hbar\right)^3\left(\beta\omega_{\rm  ho}/2\pi\right)^{3}$ with $\omega_{\rm ho}\equiv\left(\omega_x\omega_y\omega_z\right)^{1/3}$, the expression for $\gamma_0$ reduces to
\begin{eqnarray}
\gamma_0=\frac{4}{9}n\left(0\right)\left(a_2-a_0\right)^2\left(\frac{2\pi}{\beta m}\right)^{1/2}.\label{gamma_0}
\end{eqnarray}
Here, $n\left(0\right)$ is the density at the center of trap given by 
\begin{eqnarray}
n\left(0\right)=N\omega_{\rm ho}^3\left(\frac{m}{2\pi k_{\rm B}T}\right)^{3/2}.
\end{eqnarray}

We now solve the closed set of coupled moment equations. Considering the normal mode solutions $ \langle \uuline{\chi} \rangle\propto {\it e^{-i\omega t}}$, we find that Eq.(\ref{A1}) is reduced to eigenvalue equations, whose eigenvalues give frequency and damping rate $\omega=\Omega -i\gamma$.

From Eqs.(\ref{A1}) and (\ref{A2}), we find that the moment equations are divided into three individual sets of equations. First set of equations is given by
\begin{eqnarray}
\left\{
\begin{array}{l}
\omega \langle \chi_0 \rangle_{11} = -4i\gamma_0\langle \chi_0 \rangle_{-1-1}, \\
\omega \langle \chi_0 \rangle_{00} = 8i\gamma_0\langle \chi_0 \rangle_{-1-1}, \\
\omega \langle \chi_0 \rangle_{-1-1} = -4i\gamma_0\langle \chi_0 \rangle_{-1-1} .\\
\end{array}
\right. 
\end{eqnarray}
These equations describe collective mode involving oscillations of $A^{zz}$ ($A^{zz}$ mode). This mode describes purely damped motion with $\omega=-4i\gamma_0$. This means that $N$ and $M^z$ do not precession. This mode exists even in the absence of the magnetic field. The next set of equations in given by 
\begin{eqnarray}
&&\left\{
\begin{array}{l}
\displaystyle\left( \omega - \frac{\epsilon_0}{\hbar}-\frac{\epsilon_1}{\hbar}\right)\langle\chi_0\rangle_{10}+2\omega_{\rm MF}^{\left( 2 \right)} \langle\chi_0\rangle_{0-1}= 6i\gamma_0 \langle \chi_0 \rangle_{0-1} , \\
\displaystyle\left[ \omega -\frac{\epsilon_0}{\hbar}+\frac{\epsilon_1}{\hbar}-2\omega_{\rm MF}^{\left( 2 \right)}\right] \langle\chi_0\rangle_{0-1}= -6i\gamma_0 \langle \chi_0 \rangle_{0-1} ,  \\
\end{array}
\right. \\
&&\nonumber \\
&&\left\{
\begin{array}{l}
\displaystyle\left( \omega + \frac{\epsilon_0}{\hbar}+\frac{\epsilon_1}{\hbar}\right)\langle\chi_0\rangle_{01}-2\omega_{\rm MF}^{\left( 2 \right)} \langle\chi_0\rangle_{-10}= 6i\gamma_0 \langle \chi_0 \rangle_{-10} ,\\
\displaystyle\left[ \omega +\frac{\epsilon_0}{\hbar}-\frac{\epsilon_1}{\hbar}+2\omega_{\rm MF}^{\left( 2 \right)}\right] \langle\chi_0\rangle_{-10}= -6i\gamma_0 \langle \chi_0 \rangle_{-10}.  \\
\end{array}
\right. 
\end{eqnarray}
These equations describe collective modes involving oscillations of $M^\pm=\frac{1}{\sqrt{2}}\left(M^x\pm iM^y\right)$ and $A^{\pm z}=\frac{1}{\sqrt{2}}\left(A^{xz}\pm iA^{yz}\right)$ ($M^\pm$, $A^{\pm z}$ mode). Here ``$+$" mode and ``$-$" mode are degenerate in the damping rate and these frequencies. Precession of $M^\pm$ occurs only in the presence of magnetic field. On the other hand, $A^{\pm z}$ modes exist even in the absence of magnetic field. The last set of equations is given by 
\begin{eqnarray}
&&\left[ \omega -2 \frac{\epsilon_0}{\hbar}-4\omega_{\rm MF}^{\left( 2 \right)} \right]\langle \chi_0 \rangle_{1-1} =  -12i\gamma_0\langle \chi_0 \rangle_{1-1},\label{quadrupole_0_1}\\
&&\left[\omega +2 \frac{\epsilon_0}{\hbar}+4\omega_{\rm MF}^{\left( 2 \right)} \right] \langle \chi_0 \rangle_{-11} =  -12i\gamma_0\langle \chi_0 \rangle_{-11}\label{quadrupole_0_2}.
\end{eqnarray}
These equations describe collective modes involving oscillations of $A^{++}\equiv\left( A^{xx}-A^{yy}+iA^{xy}\right)/2$ and $A^{--}\equiv\left( A^{xx}-A^{yy}-iA^{xy}\right)/2$. These two modes are also degenerate in the damping rates and the frequencies have opposite sign. 

We note that the quadrupole modes exist even in the absence of external magnetic field. Although operators for conserved quantities (the total density $n$ and the magnetizations $M^\alpha$) have no values in the precession motion in these case, the quadrupole moments $A^{\alpha\beta}$ remain finite, i.e. $\langle \chi_0 \rangle\neq 0$ since they are non-conserved operators. This result is in sharp contrast with the spin-1/2 system~\cite{Nikuni2002}.

\subsection{Dipole mode} 
We next consider collective modes with dipole symmetry in a cigar-sharped trap in the presence of uniform magnetic field. To discuss these modes, we need quantities are 
\begin{eqnarray}
\chi_1 = x, \ \ \ \ \ \ \ \ \chi_2 = p_x/m .
\end{eqnarray} 
Inserting $\chi_1$ and $\chi_2$ into the general moment equation, we obtain the moment equations:
\begin{eqnarray}
\frac{d}{dt}\uuline{\langle{\chi_1}\rangle}&-&\uuline{\langle{\chi_2}\rangle} \nonumber \\
&+&\frac{i}{\hbar}\epsilon_0\uuline{\langle{\chi_1}\rangle}^A+\frac{i}{\hbar}\epsilon_1\uuline{\langle{\chi_1}\rangle}^{B} +i\omega_{\rm MF} ^{\left( 2 \right)}\uuline{\langle\chi_1\rangle}^{D}= \uuline{\langle \chi_1 \rangle}^{coll},\label{14}\\
\frac{d}{dt}\uuline{\langle{\chi_2}\rangle}&+&\omega_x^2 \uuline{\langle{\chi_1}\rangle}\nonumber \\
&+& \frac{i}{\hbar}\epsilon_0\uuline{\langle\chi_2\rangle}^A+\frac{i}{\hbar}\epsilon_1\uuline{\langle\chi_2\rangle}^{B} +i\omega_{\rm MF}^{\left( 0 \right)}\uuline{\langle\chi_2\rangle}^C+i\omega_{\rm MF} ^{\left( 2 \right)}\uuline{\langle\chi_2\rangle}^{D}= \uuline{\langle \chi_2 \rangle}^{coll},\label{15}	
\end{eqnarray}
where have introduced simplified notations for several matrices, which are defined in Appendix \ref{Appendix_Moment}.
\begin{figure}
  \begin{center}
    \scalebox{0.8}[0.8]{\includegraphics{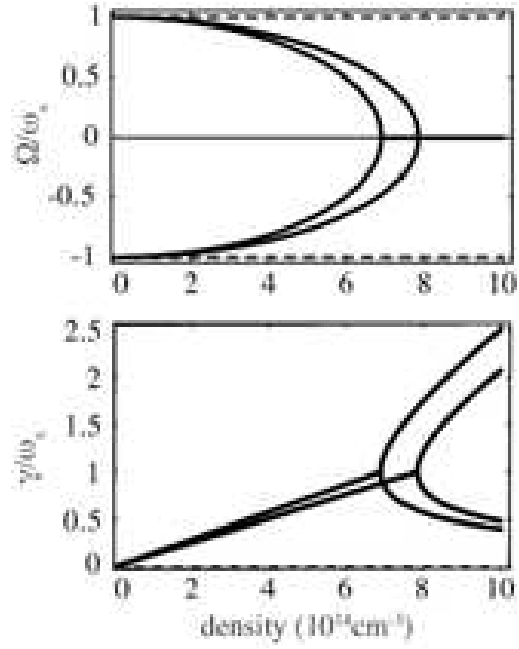}}
    \caption{Frequency and damping rate of the dipole modes obtained from Eq.(\ref{18}) versus the density at the center of the trap potential $n(0)$ for $^{23}{\rm Na}$. The dashed lines represent the dipole oscillation of the density ${\it n}$ whose frequency only depends on the trap frequency $\omega_x$. The solid lines represent the modes involving oscillations of the superposition of ${n}$, ${M^z}$, and ${A^{zz}}$.}
    \label{fig:dipole1}
  \end{center}
\end{figure}

Using the explicit form of $\delta W_{ij}$ in Eq.(\ref{13}), we obtain the collision terms as (see Appendix \ref{Appendix_Relaxation} for some details)
\begin{eqnarray}
\begin{pmatrix}
	\langle \chi_1 \rangle^{coll}_{11} & \langle \chi_1 \rangle^{coll}_{10} & \langle \chi_1 \rangle^{coll}_{1-1} \\
	\langle \chi_1 \rangle^{coll}_{01} & \langle \chi_1 \rangle^{coll}_{00} & \langle \chi_1 \rangle^{coll}_{0-1} \\
	\langle \chi_1 \rangle^{coll}_{-11} & \langle \chi_1 \rangle^{coll}_{-10} & \langle \chi_1 \rangle^{coll}_{-1-1} \\
\end{pmatrix}
= 	\begin{pmatrix}
	-2\gamma_0\langle \chi_1 \rangle_{-1-1} & 3\gamma_0\langle \chi_1 \rangle_{0-1} & -6\gamma_0\langle \chi_1 \rangle_{1-1} \\
	3\gamma_0\langle \chi_1 \rangle_{-10} &4\gamma_0\langle \chi_1 \rangle_{-1-1} & -3\gamma_0\langle \chi_1 \rangle_{0-1} \\
	-6\gamma_0\langle \chi_1 \rangle_{-11} &-3\gamma_0\langle \chi_1 \rangle_{-10} & -2\gamma_0\langle \chi_1 \rangle_{-1-1} \\
\end{pmatrix}	,	\label{16}
\end{eqnarray}
and
\begin{eqnarray}
\left\{
\begin{array}{l}
	\displaystyle\langle \chi_2 \rangle^{coll}_{11} =\Biggl.\Biggr.\frac{4}{3}\left( \gamma_2+2\gamma_1+\gamma_0 \right)\langle\chi_2\rangle_{00} +\left[ \frac{4}{3}\left(\gamma_2-2\gamma_1\right)+2\gamma_0 \right] \langle\chi_2\rangle_{-1-1} , \\
        \displaystyle\langle \chi_2 \rangle^{coll}_{10}  =\Biggl.\Biggr. -\frac{4}{3}\left( \gamma_2+2\gamma_1+\gamma_0 \right)\langle\chi_2\rangle_{10} +\frac{1}{3} \left( 8\gamma_1+5\gamma_0 \right)\langle\chi_2\rangle_{0-1} , \\
         \displaystyle\langle \chi_2 \rangle^{coll}_{1-1}  =\Biggl.\Biggr. -\frac{1}{3}\left[ 4\left(\gamma_2+2\gamma_1\right) + 22\gamma_0\right]\langle\chi_2\rangle_{1-1} , \\
	\displaystyle\langle \chi_2 \rangle^{coll}_{01} =\Biggl.\Biggr. -\frac{4}{3}\left( \gamma_2+2\gamma_1+\gamma_0\right)\langle\chi_1\rangle_{01} +\frac{1}{3}\left(8\gamma_1+5\gamma_0\right)\langle\chi_2\rangle_{-10} , \\
	\displaystyle\langle \chi_2 \rangle^{coll}_{00}  =\Biggl.\Biggr. -\frac{4}{3}\left(\gamma_2+2\gamma_1+\gamma_0\right)\langle\chi_2\rangle_{00}+4\langle\chi_2\rangle_{-1-1} , \\
	\displaystyle\langle \chi_2 \rangle^{coll}_{0-1}  =\Biggl.\Biggr. -\frac{1}{3}\left(4\gamma_2+17\gamma_0\right)\langle\chi_2\rangle_{0-1} , \\
	\displaystyle\langle \chi_2 \rangle^{coll}_{-11} =\Biggl.\Biggr. -\frac{1}{3}\left[ 4\left(\gamma_2+2\gamma_1\right) + 22\gamma_0\right]\langle\chi_2\rangle_{-11} , \\
	 \displaystyle\langle \chi_2 \rangle^{coll}_{-10} =\Biggl.\Biggr.  -\frac{1}{3}\left(4\gamma_2+17\gamma_0\right)\langle\chi_2\rangle_{-10} , \\
	 \displaystyle\langle \chi_2 \rangle^{coll}_{-1-1}=\Biggl.\Biggr. - \left[ \frac{4}{3}\left( \gamma_2-2\gamma_1\right)+6\gamma_0\right]\langle\chi_2\rangle_{-1-1} ,
\end{array}
\right. \label{17}
\end{eqnarray}
where the relaxation rate $\gamma_0$ has been defined in Appendix \ref{Appendix_Relaxation}, and $\gamma_1$ and $\gamma_2$ are defined by
\begin{eqnarray}
\left\{
\begin{array}{l}
\displaystyle\gamma_1=\frac{4}{9}n\left(0\right)\left(a_2-a_0\right)\left(a_0+2a_2\right)\left(\frac{2\pi}{\beta m}\right)^{1/2},\\
\displaystyle\gamma_2=\frac{4}{9}n\left(0\right)\left(a_0+2a_2\right)^2\left(\frac{2\pi}{\beta m}\right)^{1/2}.\\
\end{array}
\right. 
\end{eqnarray}

In summary, we obtained the coupled moment equations associated with $\chi_1=x$ and $\chi_2=p_x/m$. In addition, the collision terms couple different internal components. This implies that dipole modes do not simply describe oscillations of the density or magnetization, but oscillations of quadrupole moment are coupled to them. From Eqs.(\ref{14}), (\ref{15}), (\ref{16}), and (\ref{17}), we find that the dipole mode equations can be divided into five independent sets of equations, describing three types of modes. 

First type of modes is involves oscillations of diagonal components of moment matrices, which is described the following set of equations:
\begin{eqnarray}
\left\{
\begin{array}{l}
\displaystyle\omega \langle \chi_1 \rangle_{11} -i\langle \chi_2 \rangle_{11} = -2i\gamma_0\langle \chi_1 \rangle_{-1-1} ,\\
\displaystyle\omega \langle \chi_1 \rangle_{00} -i\langle \chi_2 \rangle_{00} = 4i\gamma_0\langle \chi_1 \rangle_{-1-1} ,\\
\displaystyle\omega \langle \chi_1 \rangle_{-1-1} -i\langle \chi_2 \rangle_{-1-1} = -2i\gamma_0\langle \chi_1 \rangle_{-1-1} ,\\
\displaystyle\omega \langle \chi_2 \rangle_{11} +i \omega_x^2 \langle \chi_1 \rangle_{11} = \frac{4i}{3} \left( \gamma_2+2\gamma_1+\gamma_0\right) \langle \chi_2 \rangle_{00} +i \left[ \frac{4}{3} \left(\gamma_2-2\gamma_1 \right) +2 \gamma_0\right] \langle \chi_2 \rangle_{-1-1} ,\\
\displaystyle\omega \langle \chi_2 \rangle_{00} + i\omega_x^2 \langle \chi_1 \rangle_{00} = -\frac{4i}{3} \left(\gamma_2+2\gamma_1+\gamma_0\right)\langle \chi_2 \rangle_{00} +4\gamma_0\langle \chi_2 \rangle_{-1-1} ,\\
\displaystyle\omega \langle \chi_2 \rangle_{-1-1} + i\omega_x^2 \langle \chi_1 \rangle_{-1-1} = - i\left[ \frac{4}{3} \left(\gamma_2-2\gamma_1\right)+6\gamma_0\right]\langle \chi_2 \rangle_{-1-1} .\\
\end{array}
\right. \label{dipole_first}
\end{eqnarray}
Second type of modes involves oscillations of the non-diagonal components, which is described by 
\begin{eqnarray}
\left\{
\begin{array}{l}
\displaystyle\left(\omega - \frac{\epsilon_0}{\hbar}- \frac{\epsilon_1}{\hbar}\right) \langle \chi_1 \rangle_{10} +\omega_{\rm MF}^{\left(2\right)}\langle\chi_1\rangle_{0-1} -i\langle \chi_2 \rangle_{10} =3i\gamma_0\langle \chi_1 \rangle_{0-1} ,\\
\displaystyle\left[\omega-\frac{\epsilon_0}{\hbar}+\frac{\epsilon_1}{\hbar}-\omega_{\rm MF}^{\left(2\right)}\right] \langle \chi_1 \rangle_{0-1} -i\langle \chi_2 \rangle_{0-1}= -3i\gamma_0\langle \chi_1 \rangle_{0-1} ,\\
\displaystyle\left[\omega-\frac{\epsilon_0}{\hbar}-\frac{\epsilon_1}{\hbar}-\omega_{\rm MF}^{\left(0\right)} -\omega_{\rm MF}^{\left(2\right)}\right]\langle \chi_2 \rangle_{10} +i \omega_x^2 \langle \chi_1 \rangle_{10}\\
\displaystyle\ \ \ \ \ \ \ \ \ \ \ \ \ \ \ \ \ \ \ \ \ \ \ \ = -\frac{4i}{3} \left( \gamma_2+2\gamma_1+\gamma_0\right) \langle \chi_2 \rangle_{10} + \frac{i}{3}\left(8\gamma_1+5\gamma_0\right)\langle \chi_2 \rangle_{0-1} ,\\
\displaystyle\left[\omega-\frac{\epsilon_0}{\hbar}+\frac{\epsilon_1}{\hbar}-2\omega_{\rm MF}^{\left(2\right)}\right] \langle \chi_2 \rangle_{0-1} + i\omega_x^2 \langle \chi_1 \rangle_{0-1}=-\frac{i}{3}\left(4\gamma_2+17\gamma_0\right)\langle \chi_2 \rangle_{0-1},\\
\end{array}
\right. \label{dipole_second_1}
\end{eqnarray}
and
\begin{eqnarray}
\left\{
\begin{array}{l}
\displaystyle\left(\omega + \frac{\epsilon_0}{\hbar}+ \frac{\epsilon_1}{\hbar}\right) \langle \chi_1 \rangle_{01} -\omega_{\rm MF}^{\left(2\right)}\langle\chi_1\rangle_{-10} -i\langle \chi_2 \rangle_{01} =3i\gamma_0\langle \chi_1 \rangle_{-10} ,\\
\displaystyle\left[\omega+\frac{\epsilon_0}{\hbar}-\frac{\epsilon_1}{\hbar}+\omega_{\rm MF}^{\left(2\right)}\right] \langle \chi_1 \rangle_{-10} -i\langle \chi_2 \rangle_{-10}= -3i\gamma_0\langle \chi_1 \rangle_{-10} ,\\
\displaystyle\left[\omega+\frac{\epsilon_0}{\hbar}+\frac{\epsilon_1}{\hbar}+\omega_{\rm MF}^{\left(0\right)} +\omega_{\rm MF}^{\left(2\right)}\right]\langle \chi_2 \rangle_{01} +i \omega_x^2 \langle \chi_1 \rangle_{01}\\
\displaystyle\ \ \ \ \ \ \ \ \ \ \ \ \ \ \ \ \ \ \ \ \ \ \ \ = -\frac{4i}{3} \left( \gamma_2+2\gamma_1+\gamma_0\right) \langle \chi_2 \rangle_{01} + \frac{i}{3}\left(8\gamma_1+5\gamma_0\right)\langle \chi_2 \rangle_{-10} ,\\
\displaystyle\left[\omega+\frac{\epsilon_0}{\hbar}-\frac{\epsilon_1}{\hbar}+2\omega_{\rm MF}^{\left(2\right)}\right] \langle \chi_2 \rangle_{-10} + i\omega_x^2 \langle \chi_1 \rangle_{-10}=-\frac{i}{3}\left(4\gamma_2+17\gamma_0\right)\langle \chi_2 \rangle_{-10} .\\
\end{array}
\right. \label{dipole_second_2}
\end{eqnarray}
Finally, the third type of modes involves oscillations of the off-diagonal component which couple $m_{\rm F}=1$ and $m_{\rm F}=-1$ states, which is described by
\begin{eqnarray}
\left\{
\begin{array}{l}
\displaystyle\left[\omega-2\frac{\epsilon_0}{\hbar}-2\omega_{\rm MF}^{\left(2\right)}\right] \langle \chi_1 \rangle_{1-1} -i\langle \chi_2 \rangle_{1-1}= -6i\gamma_0\langle \chi_1 \rangle_{1-1} ,\\
\displaystyle\left[\omega-2\frac{\epsilon_0}{\hbar}-\omega_{\rm MF}^{\left(0\right)}-3\omega_{\rm MF}^{\left(2\right)}\right] \langle \chi_2 \rangle_{1-1} +i \omega_x^2 \langle \chi_1 \rangle_{1-1}=-\frac{i}{3}\left(4\gamma_2+8\gamma_1+22\gamma_0\right) \langle \chi_2 \rangle_{1-1}  ,\\
\end{array}
\right. \label{dipole_third_1}
\end{eqnarray}
and
\begin{eqnarray}
\left\{
\begin{array}{l}
\displaystyle\left[\omega+2\frac{\epsilon_0}{\hbar}+2\omega_{\rm MF}^{\left(2\right)}\right] \langle \chi_1 \rangle_{-11} -i\langle \chi_2 \rangle_{-11}= -6i\gamma_0\langle \chi_1 \rangle_{-11} ,\\
\displaystyle\left[\omega+2\frac{\epsilon_0}{\hbar}+\omega_{\rm MF}^{\left(0\right)}+3\omega_{\rm MF}^{\left(2\right)}\right] \langle \chi_2 \rangle_{-11} +i \omega_x^2 \langle \chi_1 \rangle_{-11}=-\frac{i}{3}\left(4\gamma_2+8\gamma_1+22\gamma_0\right) \langle \chi_2 \rangle_{-11} . \\
\end{array}
\right. \label{dipole_third_2}
\end{eqnarray}

The first set of equations Eq.(\ref{dipole_first}) describes dipole oscillations of ${\it n}$, ${M^z}$ and ${A^{zz}}$. Solving the eigenvalue equation for the set of $\langle \chi_1\rangle_{11}$, $\langle \chi_1 \rangle_{00}$, $\langle \chi_1\rangle_{-1-1}$, $\langle \chi_2\rangle_{11}$, $\langle \chi_2 \rangle_{00}$, $\langle \chi_2\rangle_{-1-1}$, we obtain the complex dipole mode frequencies $\omega=\Omega-i\gamma$ as
\begin{eqnarray}
\left\{
\begin{array}{l}
\displaystyle\omega = \pm \omega_x, \\
\displaystyle\omega = -\frac{2i}{3} \left( \gamma_2+2\gamma_1+\gamma_0\right)\pm \left\{ - \left[\frac{2}{3} \left( \gamma_2+2\gamma_1+\gamma_0\right)\right]^2 + \omega_x^2 \right\}^{1/2}, \\
\displaystyle\omega = -\frac{2i}{3} \left( \gamma_2-2\gamma_1\right)-4i\gamma_0\pm \left\{-\left[\frac{2}{3}  \left( \gamma_2-2\gamma_1\right)+2\gamma_0\right]^2 + \omega_x^2 \right\}^{1/2} .
\end{array}
\right. \label{18}
\end{eqnarray}

In Fig.\ref{fig:dipole1}, we plot the frequency $\left( \Omega\right)$ and the damping rate $\left(\gamma\right)$ as a function of the density at the center of the trap potential. Here we assume ${\rm ^{23}Na}$ atoms, which have antiferromagnetic interaction. We take the following values for the physical quantities: $N=1 \times 10^5$, $\left\{\omega_x/2\pi,\omega_y/2\pi,\omega_z/2\pi \right\} = \left\{24,240,240\right\} {\rm Hz}$ and $B=20\times10^{-7} {\rm T}$. For comparison, we also calculated the frequency and the damping rate for ${\rm ^{87}Rb}$ (ferromagnetic), but we do not plot the results here, because distinction between ferromagnetic and antiferromagnetic cases in these modes are almost invisible. 

One can obtain analytical expressions for the frequencies and the damping rates of the dipole mode for Eq.(\ref{18}) in two limiting cases. For small $n\left(0\right)$, or collisionless limit $\omega \gg \left| \gamma\right|$, one has
\begin{eqnarray}
\omega\approx\pm\omega_x ,\label{19}
\end{eqnarray}
where the mode frequency only depends on the trap frequency. In general, the magnetic field does not affect these modes. For large $n\left(0\right)$, or hydrodynamic limit $\omega\ll\left| \gamma\right|$, one has
\begin{eqnarray}
\omega \approx \pm \omega_x, \ \ \ \ \ \ \omega\approx -\frac{4i}{3}\gamma_2,\ \ \ \ \ \ \omega\approx 0,\ \ \ \ \ \ \omega\approx-2i\gamma_0,\label{20}
\end{eqnarray}
where we have used the fact that $\gamma_2\gg\gamma_1\gg\gamma_0$ because $g_0\approx 200g_2$ in ${\rm ^{87}Rb}$ and $g_0\approx 30g_2$ in ${\rm ^{23}Na}$. As shown in Eq.(\ref{19}), in the collisionless limit the frequency is degenerate, only depending on the trap frequency. With increasing $n\left(0\right)$, the frequencies split and in hydrodynamic limit, reflecting that there are two kinds of modes, i.e., the mode involving fluctuations of the density and the mode involving fluctuations of $M^z$ and $A^{zz}$. The first mode, whose frequency only depends on the trap frequency, dose not decay. On the other hand, the second mode involving oscillations of $M^z$ and $A^{zz}$ are over damped. From Eq.(\ref{20}), one can see that in hydrodynamic limit only the density oscillations can be excited.

The second set of equations Eqs.({\ref{dipole_second_1}) and (\ref{dipole_second_2}) describe the dipole mode involving oscillations of ${\it M^+}$ and ${\it A^{+z}}$ as well as those of ${\it M^-}$ and ${\it A^{-z}}$. Here these two modes are degenerate in the damping rates while the frequencies have opposite sign to conserve the spin density. We only consider the mode for $M^+$ and $A^{+z}$. Solving the eigenvalue equation for the set of variables $\langle \chi_1\rangle_{10}$, $\langle \chi_1 \rangle_{0-1}$, $\langle \chi_2\rangle_{10}$, $\langle \chi_2 \rangle_{0-1}$, we obtain the dipole mode frequencies $\omega=\Omega-i\gamma$ as
\begin{eqnarray}
\left\{
\begin{array}{l}
 \displaystyle\omega = \frac{1}{2} \Biggl(\frac{2}{\hbar}\left(\epsilon_0+\epsilon_1\right)+ \omega_{\rm MF}^{\left( 0 \right)} +\omega_{\rm MF}^{\left( 2 \right)}- \frac{4i}{3} \left( \gamma_2+2\gamma_1+\gamma_0 \right) \Biggr.\\
 \ \ \ \ \ \ \ \ \ \ \ \ \ \ \ \ \ \ \ \ \ \ \ \ \ \ \ \ \ \ \ \ \ \ \ \ \ \ \Biggl.
  \pm \left\{ \left[\omega_{\rm MF}^{\left( 0 \right)} + \omega_{\rm MF}^{\left( 2 \right)} -\frac{4i}{3} \left(\gamma_2+2\gamma_1+\gamma_0\right)\right]^2 + 4\omega_x^2 \right\}^{1/2}\Biggr) ,\\
 \displaystyle \omega = \frac{1}{2} \left(\frac{2}{\hbar}\left(\epsilon_0-\epsilon_1\right)+3\omega_{\rm MF}^{\left( 2 \right)}-\frac{4i}{3}\gamma_2-\frac{26i}{3}\gamma_0 
  \pm \left\{ \left[\omega_{\rm MF}^{\left( 2 \right)} -\frac{4i}{3} \left(\gamma_2+2\gamma_0\right)\right]^2 + 4\omega_x^2 \right\}^{1/2}\right).
\end{array}
\right. \label{21}
\end{eqnarray}
In Fig.\ref{fig:dipole2},  we plot the frequency $\left(\Omega\right)$ and damping rate $\left(\gamma\right)$ for both ${\rm ^{87}Rb}$ and ${\rm ^{23}Na}$. From Fig.\ref{fig:dipole2}, one clearly sees the distinction between ferromagnetic or antiferromagnetic cases. 
\begin{figure}
  \begin{center}
    \begin{tabular}{cc}
      \scalebox{0.8}[0.8]{\includegraphics{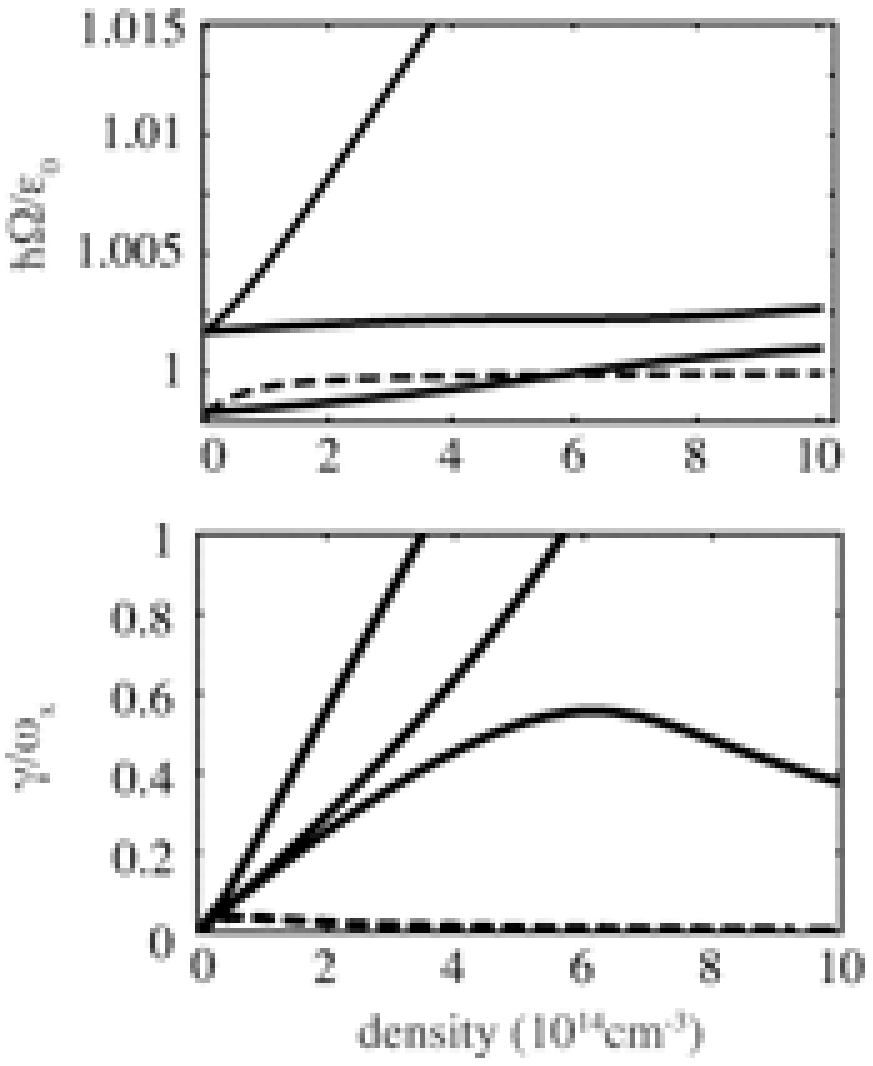}} &
      \scalebox{0.8}[0.8]{\includegraphics{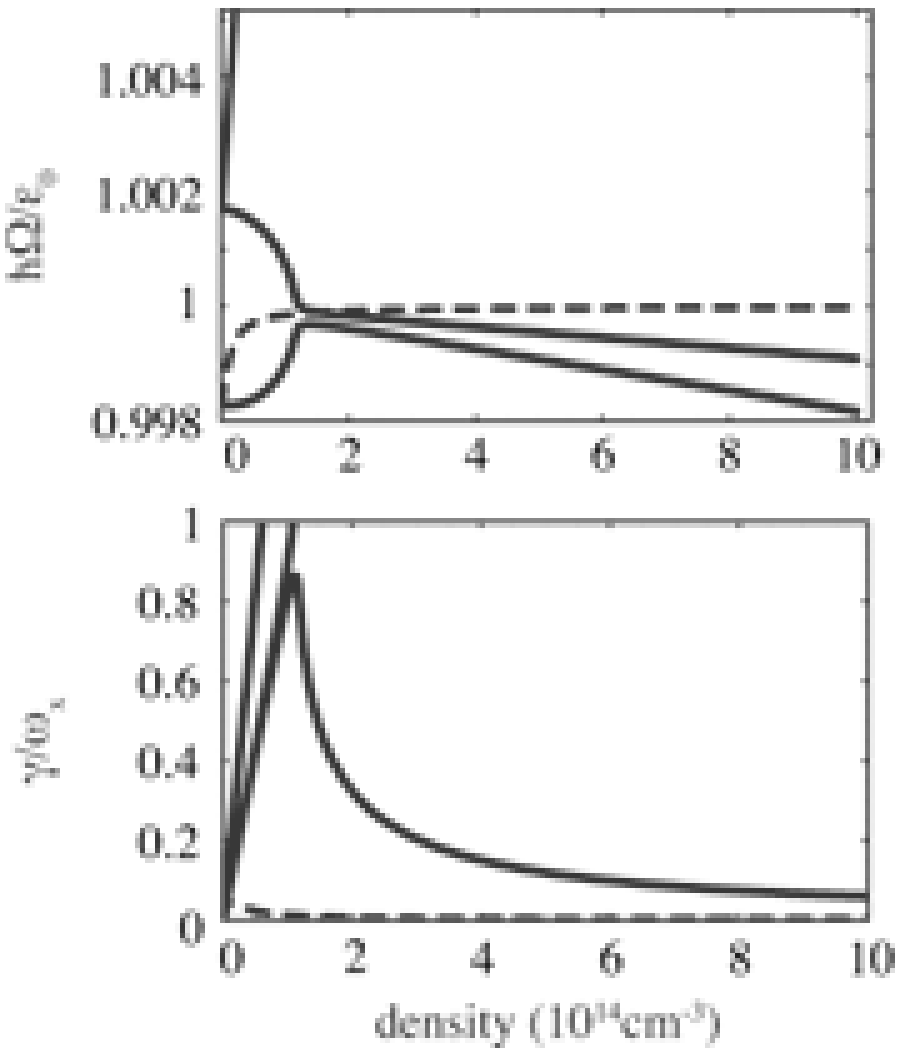}}
    \end{tabular}
    \caption{Frequency and damping rate of the dipole mode obtained from Eq.(\ref{21}) versus the density at the center of the trap potential $n(0)$ for ${\rm ^{23}Na}$ (left) and ${\rm ^{87}Rb}$ (right). The dashed lines represent the modes having the lowest damping rates among the four modes. The solid lines represent the other dipole modes.}
    \label{fig:dipole2}
  \end{center}
\end{figure}

In the collisionless limit, the solutions of Eq.(\ref{21}) reduces to
\begin{eqnarray}
\omega \approx \frac{1}{\hbar}\left(\epsilon_0 + \epsilon_1\right) \pm\omega_x, \ \ \ \ \ \ \ \  \omega \approx \frac{1}{\hbar}\left(\epsilon_0 - \epsilon_1\right) \pm\omega_x.\label{26}
\end{eqnarray} 
Corresponding eigenvectors are given by
\begin{eqnarray}
\begin{pmatrix}
\langle\chi_1\rangle_{10}\\
\langle\chi_1\rangle_{0-1}\\
\langle\chi_2\rangle_{10}\\
\langle\chi_2\rangle_{0-1}
\end{pmatrix}
=\begin{pmatrix}
0\\
\pm\frac{1}{\omega_x}\\
0\\
i
\end{pmatrix}, \ \ \ \ \ \ \ \  
\begin{pmatrix}
\langle\chi_1\rangle_{10}\\
\langle\chi_1\rangle_{0-1}\\
\langle\chi_2\rangle_{10}\\
\langle\chi_2\rangle_{0-1}
\end{pmatrix}
=\begin{pmatrix}
\pm\frac{1}{\omega_x}\\
0\\
i\\
0
\end{pmatrix}.
\end{eqnarray}
As expected, the quadratic Zeeman energy splits the frequencies of two modes, and the frequencies strongly depend on the magnetic field in this collisionless limit. 

In the hydrodynamic limit, one has
\begin{eqnarray}
\omega \approx 0, \ \ \ \  \omega\approx\-\left(\omega_{\rm MF}^{\left(0\right)}+\omega_{\rm MF}^{\left(2\right)}\right)-\frac{4i}{3}\gamma_2, \ \ \ \  \omega\approx \omega_{\rm MF}^{\left(2\right)}-3i\gamma_0, \ \ \ \ \omega\approx2\omega_{\rm MF}^{\left(0\right)}-\frac{4i}{3}\gamma_2,\label{22}
\end{eqnarray}
where we made use of the fact $\gamma_2\gg\gamma_1\gg\gamma_0$, since we now consider ${\rm ^{87}Rb} $ and ${\rm ^{23}Na} \left( a_2\simeq a_0\right)$. The last two solutions in Eq.(\ref{22}) are strongly affected by the sign of $g_2$. 
\begin{figure}
  \begin{center}
    \scalebox{1.0}[1.0]{\includegraphics{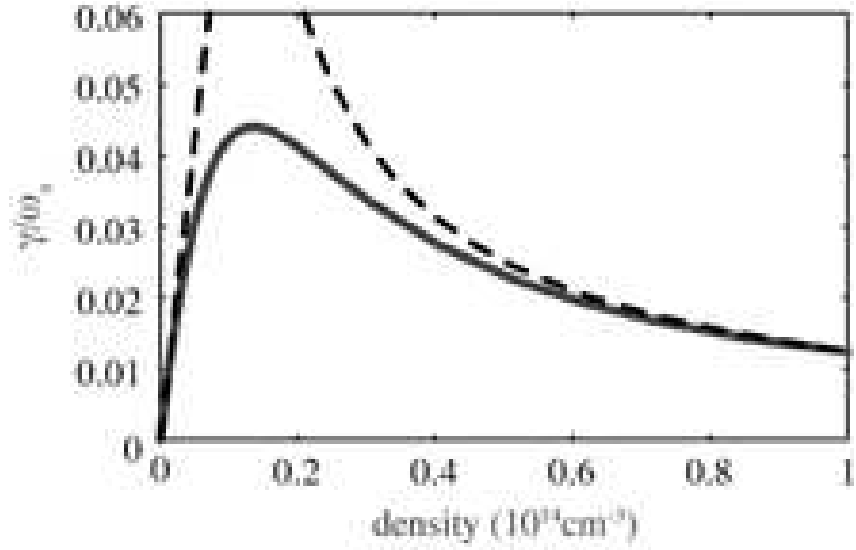}}
    \caption{The damping rate of the mode with the lowest damping rate in Fig.\ref{fig:dipole2} versus the density at the center of the trap potential $n(0)$ for ${\rm ^{87}Rb}$. The solid line is the damping rate obtain from Eq.(\ref{21}) and the dashed lines are in the collisionless limit Eq.(\ref{23}) and the hydrodynamic limit Eq.(\ref{24}), respectively. }
    \label{fig:dampingrate}
  \end{center}
\end{figure}

In Fig.\ref{fig:dampingrate}, we focus on the mode with the lowest damping rate. The frequency of this mode grows linearly with $n\left(0\right)$, while the damping rate has a peak. This kind of peak behavior of damping rate is often seen in cross over from collisionless to hydrodynamic modes~\cite{Baym_book,Nikuni2002}.

In the collisionless limit, the damping rate grows as
\begin{eqnarray}
\Delta\omega\approx\frac{2}{3}\gamma_2,\label{23}
\end{eqnarray}     
while in the hydrodynamic limit, the damping rate falls as
\begin{eqnarray}
\Delta\omega\approx -\frac{12\omega_x^2 \gamma_2}{\left(4\gamma_2\right)^2+9\left(\omega_{\rm MF}^{\left(0\right)}+\omega_{\rm MF}^{\left(2\right)}\right)^2}.\label{24}
\end{eqnarray}

\begin{figure}
  \begin{center}
    \scalebox{0.8}[0.8]{\includegraphics{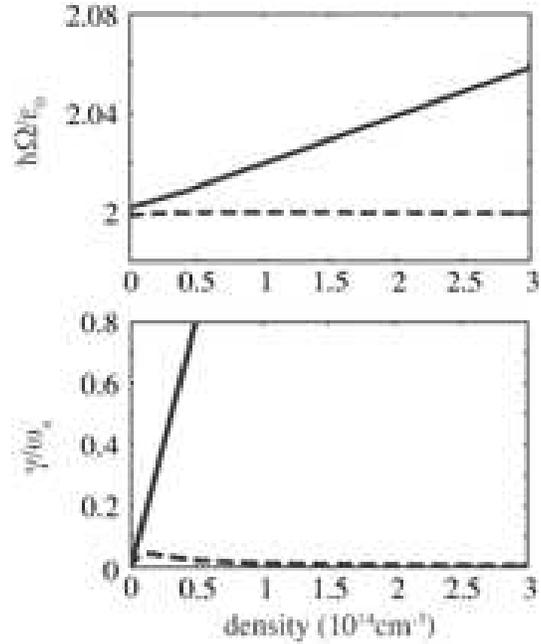}}
    \caption{Frequency and damping rate of the dipole mode obtained from Eq.(\ref{25}) versus the density at the center of the trap potential $n(0)$ for ${\rm ^{87}Rb}$. The dashed lines represent the mode having the lowest damping rate. The solid lines represent the other mode.}
    \label{fig:dipole4}
  \end{center}
\end{figure}
In Fig.\ref{fig:dampingrate}, we also plot these asymptotic behaviors given by Eqs.(\ref{23}) and (\ref{24}). Moreover, we consider how oscillations of $M^+$ and $A^{+z}$ are coupled. Calculating the eigenvalue equation to find the eigenvector, we find that this mode involves oscillations of both $M^+$ and $A^{+z}$ in the collisionless limit. While in the hydrodynamic limit, the non-conserving quantity $A^{+z}$ decays quickly, so this mode only involve of oscillations $M^{+}$.

The third set of equations Eqs.(\ref{dipole_third_1}) and (\ref{dipole_third_2}) describe dipole modes  involving oscillations of $A^{++}$ as well as $A^{--}$, which couple $m_{\rm F}=1$ state with $m_{\rm F}=-1$. These modes involve the superposition of quadrupole moments which is constructed by multiplying $M^x$ and $M^y$. These two modes are also degenerate in the damping rates and the frequencies have opposite sign. We thus only consider the modes for $A^{++}$.
Solving the eigenvalue equation for the set of $\langle \chi_1\rangle_{1-1}$, $\langle \chi_2 \rangle_{1-1}$, we obtain the dipole mode frequencies $\omega=\Omega-i\gamma$ as
\begin{eqnarray}
 \omega &=& \frac{1}{2} \Biggl(\frac{4}{\hbar}\epsilon_0+\omega_{\rm MF}^{\left( 0 \right)} + 5\omega_{\rm MF}^{\left( 2 \right)} - \left[\frac{4i}{3} \left( \gamma_2+2\gamma_1+\gamma_0 \right)+12\gamma_0\right] \Biggr. \nonumber \\
 & & \ \ \ \ \ \ \Biggl. \pm \left\{ \left[\omega_{\rm MF}^{\left( 0 \right)}+\omega_{\rm MF}^{\left( 2 \right)} -\frac{4i}{3} \left( \gamma_2+2\gamma_1+\gamma_0\right)\right]^2 + 4\omega_x^2 \right\}^{1/2}\Biggr) .\label{25}
\end{eqnarray}
From Eq.(\ref{25}), one sees that the oscillations are independent of the quadratic Zeeman energy. In Fig.\ref{fig:dipole4}, we plot the frequency $\left(\Omega\right)$ and the damping rate $\left(\gamma\right)$ given from Eq.(\ref{25}).
In the collisionless limit, the frequency strongly depends on the magnetic field as
\begin{eqnarray}
\omega\approx\frac{2}{\hbar}\epsilon_0\pm\omega_x. \label{27}
\end{eqnarray}
In the hydrodynamic limit, the mean-field strongly affects the frequency, which can be approximated as
\begin{eqnarray}
\omega\approx2\omega_{\rm MF}^{\left(2\right)}-6i\gamma_0, \ \ \ \ \ \ \ \  \omega\approx\omega_{\rm MF}^{\left(0\right)}+3\omega_{\rm MF}^{\left(2\right)}-\frac{4i}{3}\gamma_2.
\end{eqnarray}
 \begin{figure}
  \begin{center}
    \scalebox{1.0}[1.0]{\includegraphics{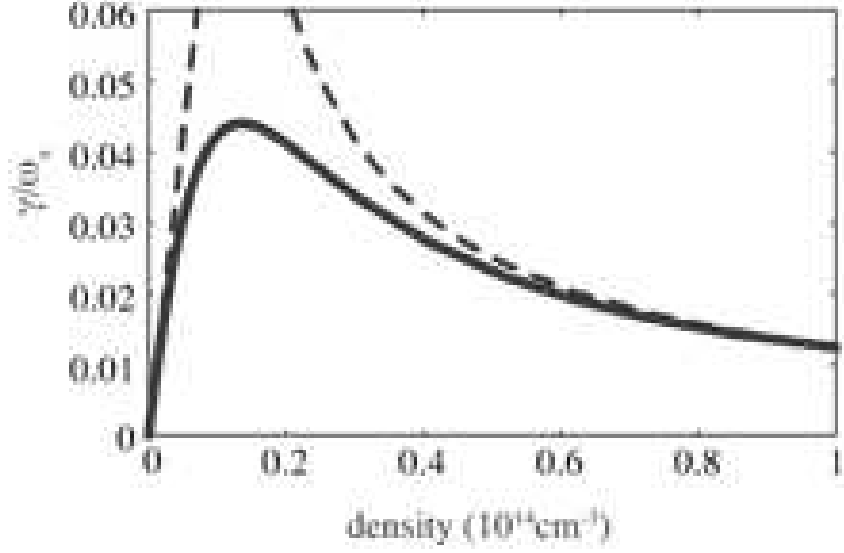}}
    \caption{The damping rate of the mode with the lowest damping rate versus the density at the center of the trap potential $n(0)$ at for ${\rm ^{87}Rb}$. The solid lime is the damping rate obtained from Eq.(\ref{25}), and the dashed lines represent the  asymptotic behaviors in the collisionless limit and that of the hydrodynamic limit respectively. }
    \label{fig:dampingrate2}
  \end{center}
\end{figure}
We see that oscillations of the quadrupole moments cannot be easily excited. As shown in Fig.\ref{fig:dipole4}, one of the two modes has a high damping rate, which implies that this mode decays quickly. The both modes have high frequencies, which imply that these modes cannot be easily excited. We now focus on the oscillation with the lowest damping rate. In Fig.\ref{fig:dampingrate2}, we plot the lowest damping rate as well as its asymptotic behaviors. Similarly to Fig.\ref{fig:dampingrate}, we find a peak behavior.

Finally, we briefly discuss experimental possibility of observing spin-wave collective modes predicted in the present paper. In the JILA experiment on the spin-wave collective modes of the spin-1/2 system~\cite{McGuirk2002}, quadrupole mode is excited by quadratically varying frequency splitting $\Delta\left(\textbf{r}\right)\propto z^2$, where $\Delta\left(\textbf{r}\right)$ is the transition frequency between the two states, which is equivalent to the transverse magnetic field in our spin-1 system. In this experiment, frequency and damping rate of the spin-wave oscillation was investigated in great detail. In addition to the quadrupole mode, the dipole mode of the spin-1/2 system is expected to be excited by a linear inhomogeneous frequency splitting, such as $\Delta\left(\textbf{r}\right)\propto z$~\cite{Nikuni2002}. Analogously to the previous studies on the spin-1/2 system, the dipole mode of the spin-1 system can be excited by using linear field gradient.  We note that the linear Zeeman field $B^\alpha$ will excite the collective mode involving oscillations of the linear magnetization $M^\alpha$. In order to excite the modes dominated by the quadrupole moment $A^{\alpha\beta}$, one needs to apply quadrupole coupling field $B^{\alpha\beta}$, that couples $m_F=+1$ state and $m_F=-1$ state. Such an external field may be created by a two-photon coupling field.

\section{Conclusions} 
In this paper, we have studied the collective oscillation in a dilute noncondensed Bose gas in an optical trap. First, we derived the kinetic equation describing the nonequilibrium behavior of the system to discuss the spin-1 Bose gas at finite temperatures. Starting from this equation, we derived the moment equation for the dipole mode given in Eqs.(\ref{14}) and (\ref{15}). Solving those equations, we found the following three properties of the dipole mode. First, the mode can be classified into three independent modes; the first kind of modes involves oscillations for the density {\it n}, {\it z}-component of magnetization $M^z$ and {\it zz}-component of quadrupole moment $A^{zz}$, the second kind of modes involve oscillations of the superposition of $M^+$ and $A^{+z}$ and that of $M^-$ and $A^{-z}$, and the third kind of modes involve oscillations of $A^{++}$ and that of $A^{--}$. Second, we found that the damping rate is characterized by the three relaxation rates. These relaxation rates are expressed as linear functions of the density $n\left(0\right)$ at the center of the trap potential. Finally, we considered the frequency and damping rate of the dipole mode. It is very interesting that the spin-1 thermal gas exhibits strong collective behavior due to the spin-spin interaction even in the collisionless regime. Moreover, we found that $A^{\pm z}$ and $A^{\pm\pm}$ modes can exist even in the absence of magnetic field, as shown in Eqs.(\ref{quadrupole_0_1}), (\ref {quadrupole_0_2}), and (\ref{25}). We expect this type of  quadrupole modes to be excited by using a two-photon coupling field. 

In this paper, we only considered the normal gas above the BEC transition temperature $T_{\rm BEC}$.
In the Bose-condensed phase below $T_{\rm BEC}$, the spin of condensate and thermal cloud gas will interact strongly~\cite{Erhard2004,Schmaljohann2004}. For future study, we will investigate behaviors of the spin-1 Bose gas including condensate, and study how the condensate dynamics is affected by the thermal component.

\appendix

\section{Collision integrals}\label{Appendix_Collision}
In this appendix, we calculate the collision term Eq.(\ref{7}) explicitly. Following Refs.~\cite{Zaremba1999, Nikuni2003}, we assume that the effect of $H^\prime\left( t\right)$ is essentially a collision process, which occurs on a time scale much shorter than all other time scales in the problem. We also assume that the hydrodynamic variables vary slowly in space and time. With these two key assumptions, one therefore expects that the dominant contribution from the commutator of field operators $\hat\Psi_i\left(\textbf{r},t\right)$ and terms contained in the effective Hamiltonian involved in Eq.(\ref{7}) to arise from values of $\textbf{r}^\prime$ and $t^\prime$ close to $\textbf{r}$ and $t$. In this situation, we can expand the quantites $n_{ij}$ and $\uuline{U}$ that appear in $\hat H$ about this point. It is sufficient to use  
\begin{eqnarray}
n_{ij} \left( \textbf{r}^\prime,t^\prime \right)  \approx n_{ij} \left( \textbf{r},t \right),\ \ \ \ \uuline{U}\left(\textbf{r}^\prime,t^\prime\right)\approx \uuline{U}\left(\textbf{r},t\right).
\end{eqnarray}
With this approximation, $\hat H_{\rm MF}$ and $\hat H^\prime$ are given as 
\begin{eqnarray}
\hat H_{\rm MF}&=& \sum_{ii^\prime jj^\prime} V_{ii^\prime jj^\prime} \left[  n_{j^\prime i}\left( \textbf{r},t \right)\int d\textbf{r}^\prime \hat \Psi ^\dagger _{i^\prime} \left( \textbf{r}^\prime \right) \hat \Psi ^\dagger _{j} \left( \textbf{r}^\prime \right) + n_{ji}\left( \textbf{r},t \right)\int d\textbf{r}^\prime \hat \Psi ^\dagger _{i^\prime} \left( \textbf{r}^\prime  \right) \hat \Psi ^\dagger _{j^\prime} \left( \textbf{r}^\prime \right)  \right],\\
\hat H^\prime  &=& \frac{1}{2} \sum_{ii^\prime jj^\prime} V_{ii^\prime jj^\prime}\int d\textbf{r}                         \hat \Psi ^\dagger _{i} \left( \textbf{r} \right) \hat \Psi ^\dagger _{i^\prime} \left( \textbf{r} \right)\hat \Psi _{j^\prime} \left( \textbf{r} \right) \hat \Psi _{j} \left( \textbf{r} \right)  \nonumber \\ 
                   & & - \sum_{ii^\prime jj^\prime} V_{ii^\prime jj^\prime} \left[  n_{j^\prime i}\left( \textbf{r},t \right)\int d\textbf{r}^\prime \hat \Psi ^\dagger _{i^\prime} \left( \textbf{r}^\prime \right) \hat \Psi ^\dagger _{j} \left( \textbf{r}^\prime \right) + n_{ji}\left( \textbf{r},t \right)\int d\textbf{r}^\prime \hat \Psi ^\dagger _{i^\prime} \left( \textbf{r}^\prime  \right) \hat \Psi ^\dagger _{j^\prime} \left( \textbf{r}^\prime \right)  \right] .
\end{eqnarray}
It is now useful to introduce the Fourier transform of the field operator
\begin{eqnarray}
\hat \Psi _i \left( \textbf{r},t \right) = \frac{1}{\sqrt{V}} \sum_{\textbf{p}} a_{i \textbf{p}}e^{i\textbf{p} \cdot \textbf{r} / \hbar },
\end{eqnarray}
where $V$ is the volume of the system. One can then write $\hat H_{\rm MF}$ and $\hat H^\prime$ as
\begin{eqnarray}
\hat H_{\rm MF}&=&\sum_{ii^\prime jj^\prime} \sum_{\textbf{p}}V_{ii^\prime jj^\prime}\left[ n_{ji}\left( \textbf{r},t \right)a_{i^\prime \textbf{p}}^\dagger a_{j^\prime \textbf{p}} + n_{j^\prime i}\left( \textbf{r},t \right)a_{i^\prime \textbf{p}}^\dagger a_{j \textbf{p}} \right] , \\
\hat H^\prime  &=& \frac{1}{V}\sum_{ii^\prime jj^\prime} \sum_{\textbf{p}_1\textbf{p}_2\textbf{p}_3\textbf{p}_4}V_{ii^\prime jj^\prime}\delta_{\textbf{p}_1 + \textbf{p}_2 , \textbf{p}_3 + \textbf{p}_4 }a_{i\textbf{p}_1}^\dagger a_{i^\prime\textbf{p}_2}^\dagger a_{j^\prime \textbf{p}_3}a_{j \textbf{p}_4} \nonumber \\
              & & -\sum_{ii^\prime jj^\prime} \sum_{\textbf{p}}V_{ii^\prime jj^\prime}\left[ n_{ji}\left( \textbf{r},t \right)a_{i^\prime \textbf{p}}^\dagger a_{j^\prime \textbf{p}} + n_{j^\prime i}\left( \textbf{r},t \right)a_{i^\prime \textbf{p}}^\dagger a_{j \textbf{p}} \right] \label{6}.
\end{eqnarray}
It should be understood that the above ``local'' approximation for the effective Hamiltonian can only be used in calculating the collision integral at specific position and time $\left(\textbf{r},t\right)$.

Using the time-dependent perturbation theory, one can expand a statistical average of an operator $\hat O$ as
\begin{eqnarray}
\langle \hat O \rangle_t &=& {\rm tr} \hat \rho \left( t_0 \right) \Biggl\{ {\cal U}_0^\dagger \left( t,t_0 \right) \hat O \left( t_0 \right) \hat {\cal U}_0 \left( t,t_0 \right) \Biggr. \nonumber \\
                         & & \Biggl. \ \ \ \ \ \ \ \ \ \ \ \ -\frac{i}{\hbar }\int_{t_0}^{t}dt^\prime \hat {\cal U}_0 \left( t^\prime ,t_0 \right) \left[ \hat {\cal U}_0^\dagger  \left( t,t^\prime \right) \hat O \left( t_0 \right) \hat {\cal U}_0 \left( t,t^\prime \right) , \hat H^\prime \left( t^\prime \right) \right] \hat {\cal U}_0  \left( t^\prime,t_0 \right) \Biggr\} , 
\end{eqnarray}
where we have introduced the unperturbed evolution operator
\begin{eqnarray}
\hat{\cal U}_0 \left( t,t_0 \right) = {\it T}  \exp \left[ - \frac{i}{\hbar }\int_{t_0}^{t} dt^\prime \hat H_{\rm MF} \left( t^\prime \right) \right],
\end{eqnarray}
with `` {\it T} '' being the time-ordering operator. Following Refs.~\cite{Zaremba1999,Nikuni2003}, we use the simple approximation for a free evolution of $a_{i\textbf{p}}$ 
\begin{eqnarray}
\hat {\cal U}_0^\dagger  \left( t,t^\prime \right) a_{i\textbf{p}} \hat {\cal U}_0 \left( t,t^\prime \right)%
     \approx \exp \left[ - \frac{i}{\hbar } \epsilon _p \left( t - t^\prime \right) \right]a_{i\textbf{p}},
\end{eqnarray}
where $\epsilon_p \equiv p^2/2m$. As usual, we assume that there is no initial correlation so that we can use the Wick's theorem to factorize terms such as
\begin{eqnarray}
\langle a_{i\textbf{p}_1}^\dagger a_{j\textbf{p}_2}^\dagger a_{k\textbf{p}_3} a_{l\textbf{p}_4}\rangle_{t_0} = \langle a_{i\textbf{p}_1}^\dagger a_{k\textbf{p}_3}\rangle_{t_0}\langle a_{j\textbf{p}_2}^\dagger a_{l\textbf{p}_4}\rangle_{t_0} +\langle a_{i\textbf{p}_1}^\dagger a_{l\textbf{p}_4}\rangle_{t_0}\langle a_{j\textbf{p}_2}^\dagger a_{k\textbf{p}_3}\rangle_{t_0} .
\end{eqnarray}
Within the present assumption of slow variation of the macroscopic quantities in space and time, we can use 
\begin{eqnarray}
\langle a_{i \textbf{p}_1}^\dagger a_{j \textbf{p}_2} \rangle_{t_0} \approx \delta _{\textbf{p}_1,\textbf{p}_2} W_{ij} \left( \textbf{p},\textbf{r},t \right),
\end{eqnarray}
in the calculation of the collision integral. 

Using these approximations, we obtain the following contribution to the collision term Eq.(\ref{7}) as
\begin{eqnarray}
&&{\rm tr} \hat \rho \left( t_0 \right) \left[ \hat W_{ij}\left( \textbf{p},\textbf{r},t \right) ,\hat H^\prime \left( t \right) \right]\nonumber \\
&& \ \ \ \ \ \ = -\frac{i}{\hbar V^2}\sum_{j^{\prime\prime}}\sum_{\textbf{p}_1\textbf{p}_2\textbf{p}_3}\delta _{\textbf{p} + \textbf{p}_1,\textbf{p}_2 + \textbf{p}_3}\int_{t_0}^{t}dt^\prime  e^{i\left(\epsilon_{p} + \epsilon_{p_1} - \epsilon_{p_2} - \epsilon_{p_3} \right) \left( t-t^\prime \right)/\hbar } \nonumber \\
  & & \ \ \ \ \ \ \ \ \ \ \ \ \ \ \times \left[ \Sigma_{ij^{\prime\prime}}^> \left( \textbf{p}_1,\textbf{p}_2,\textbf{p}_3 \right)W_{j^{\prime\prime}j}^< \left( \textbf{p} \right) -\Sigma_{ij^{\prime\prime}}^< \left( \textbf{p}_1,\textbf{p}_2,\textbf{p}_3 \right)W_{j^{\prime\prime}j}^> \left( \textbf{p} \right)  \right] \nonumber \\
&& \ \ \ \ \ \ +\frac{i}{\hbar V^2}\sum_{i^{\prime\prime}}\sum_{\textbf{p}_1\textbf{p}_2\textbf{p}_3}\delta _{\textbf{p} + \textbf{p}_1,\textbf{p}_2 + \textbf{p}_3}\int_{t_0}^{t}dt^\prime e^{-i\left(\epsilon_{p} + \epsilon_{p_1} - \epsilon_{p_2} - \epsilon_{p_3} \right) \left( t-t^\prime \right)/\hbar } \nonumber \\
  & & \ \ \ \ \ \ \ \ \ \ \ \ \ \ \times \left[ \Sigma_{i^{\prime\prime}j}^< \left( \textbf{p}_1,\textbf{p}_2,\textbf{p}_3 \right)W_{ii^{\prime\prime}}^> \left( \textbf{p} \right) -\Sigma_{i^{\prime\prime}j}^> \left( \textbf{p}_1,\textbf{p}_2,\textbf{p}_3 \right)W_{ii^{\prime\prime}}^< \left( \textbf{p} \right)  \right] , \label{8}
\end{eqnarray}
where we have introduced the following notations:
\begin{eqnarray}
W_{ij}^< \left( \textbf{p} \right) \equiv W_{ij} \left( \textbf{p} \right),\ \ \ \ W_{ij}^> \left( \textbf{p} \right) \equiv \delta _{ij} + W_{ij} \left( \textbf{p} \right), 
\end{eqnarray}
\begin{eqnarray}
&& \Sigma_{ij^{\prime\prime}}^{{< \atop >}} \left( \textbf{p}_1,\textbf{p}_2,\textbf{p}_3 \right)\equiv \sum_{ll^\prime mi^\prime i^{\prime\prime}j^\prime}V_{imll^\prime}V_{i^\prime i^{\prime\prime}j^\prime j^{\prime\prime}}\left[W_{j^\prime m}^{> \atop <}\left( \textbf{p}_1 \right)W_{l^\prime i^\prime}^{< \atop >}\left( \textbf{p}_2 \right) W_{li^{\prime\prime}}^{< \atop >}\left( \textbf{p}_3 \right) \right.\nonumber \\
&& \ \ \ \ \ \ \ \ \ \ \ \ \ \ \ \ \ \ \ \ \ \ \ \ \ \ \ \ \ \ \ \ \ \ \ \ \ \ \ \ \ \ \ \ \ \ \ \ \ \ \left.+ W_{j^\prime m}^{> \atop <}\left( \textbf{p}_1 \right)W_{l^\prime i^{\prime\prime}}^{< \atop >}\left( \textbf{p}_2 \right)W_{li^{\prime}}^{< \atop >}\left( \textbf{p}_3 \right)\right] ,
\\
&& \Sigma_{i^{\prime\prime}j}^{{< \atop >}} \left( \textbf{p}_1,\textbf{p}_2,\textbf{p}_3 \right) \equiv \sum_{ll^\prime mi^\prime j^\prime j^{\prime\prime}}V_{ll^\prime mj}V_{i^\prime i^{\prime\prime}j^\prime j^{\prime\prime}}\left[W_{mi^\prime}^{> \atop <}\left( \textbf{p}_1 \right)W_{j^\prime l^\prime}^{< \atop >}\left( \textbf{p}_2 \right)W_{j^{\prime\prime}l}^{< \atop >}\left( \textbf{p}_3 \right) \right.\nonumber \\
&& \ \ \ \ \ \ \ \ \ \ \ \ \ \ \ \ \ \ \ \ \ \ \ \ \ \ \ \ \ \ \ \ \ \ \ \ \ \ \ \ \ \ \ \ \ \ \ \ \ \ \left.+ W_{mi^\prime}^{> \atop <}\left( \textbf{p}_1 \right)W_{j^{\prime\prime}l^\prime}^{< \atop >}\left( \textbf{p}_2 \right)W_{j^{\prime}l}^{< \atop >}\left( \textbf{p}_3 \right) \right],
\end{eqnarray}  
omitted the arguments $\textbf{r}$ and $t$ for simplicity. In the time integral of Eq.(\ref{8}), we set $t_0 \rightarrow -\infty$, which yields
\begin{eqnarray}
&& \frac{1}{\hbar }\int_{-\infty }^{t}dt^\prime e^{i\left(\epsilon_{p} + \epsilon_{p_1} - \epsilon_{p_2} - \epsilon_{p_3} \right) \left( t-t^\prime \right)/\hbar } \nonumber \\
&& \ \ \ \ \ \ \ \ =\pi \delta \left( \epsilon_p + \epsilon_{p_1} - \epsilon_{p_2} - \epsilon_{p_3} \right)%
     + i{\cal P} \left( \frac{1}{\epsilon_p + \epsilon_{p_1} - \epsilon_{p_2} - \epsilon_{p_3}} \right),  
\end{eqnarray}
where $ {\cal P}$ is the symbol of principal integration. We then obtain 
\begin{eqnarray}
\uuline{I} &=& \left.\frac{\partial \uuline{W}}{\partial t}\right|_{coll} + \frac{i}{\hbar }\left[ \uuline{W}\left(\textbf{p}\right),\delta\uuline{U_n} \left( \textbf{p} \right) \right] ,
\end{eqnarray}
where $\frac{\partial \uuline{W}}{\partial t}\mid_{coll}$ is the collision integral and $\delta \uuline{U_n}$ is the second order potential given by
\begin{eqnarray}
&&\left.\frac{\partial \uuline{W}}{\partial t}\right|_{coll} \equiv \frac{\pi }{\hbar V^2} \sum_{\textbf{p}_1\textbf{p}_2\textbf{p}_3} \delta \left( \epsilon_p + \epsilon_{p_1} - \epsilon_{p_2} - \epsilon_{p_3} \right) \delta _{\textbf{p} + \textbf{p}_1,\textbf{p}_2 + \textbf{p}_3} \nonumber \\
&& \ \ \ \ \ \ \ \ \ \ \ \ \ \ \ \ \ \ \ \ \times \left[ \left\{ \uuline{1} + \uuline{W} \left( \textbf{p}\right) ,\uuline{\Sigma}^< \left( \textbf{p}_1,\textbf{p}_2,\textbf{p}_3 \right) \right\} - \left\{ \uuline{W} \left( \textbf{p}\right) ,\uuline{\Sigma}^> \left( \textbf{p}_1,\textbf{p}_2,\textbf{p}_3 \right)\right\} \right] ,\label{coll2}\\ 
&& \delta \uuline{U_n} \left( \textbf{p} \right) \equiv \frac{1}{V^2}\sum_{\textbf{p}_1\textbf{p}_2\textbf{p}_3}{\cal P} \left( \frac{1}{\epsilon_p + \epsilon_{p_1} - \epsilon_{p_2} - \epsilon_{p_3}}\right) \delta _{\textbf{p} + \textbf{p}_1,\textbf{p}_2 + \textbf{p}_3} \nonumber \\
&& \ \ \ \ \ \ \ \ \ \ \ \ \ \ \ \ \ \ \times \left[ \uuline{\Sigma}^> \left( \textbf{p}_1,\textbf{p}_2,\textbf{p}_3 \right) - \uuline{\Sigma}^< \left( \textbf{p}_1,\textbf{p}_2,\textbf{p}_3 \right) \right].\label{coll}
\end{eqnarray}
For a dilute Bose gas considered in this paper, one can neglect $\delta \uuline{U_n}$~\cite{Zaremba1999,Nikuni2003}. Finally, we replace the momentum sum $1/V\sum_{\textbf{p}}$ by the integral $\int d\textbf{p}/\left(2\pi\hbar\right)^3$ and the Kronecker delta function $V\delta_{\textbf{p},\textbf{p}^\prime}$ by the Dirac delta function $\left(2\pi\hbar\right)^3\delta\left(\textbf{p}-\textbf{p}^\prime\right)$. We then obtain the expression for the collision integral
\begin{eqnarray}
&&\left.\frac{\partial \uuline{W}}{\partial t}\right|_{coll} \equiv \frac{\pi }{\hbar}\int d\textbf{p}_1\int \frac{d\textbf{p}_2}{\left(2\pi\hbar\right)^3}\int \frac{d\textbf{p}_3}{\left(2\pi\hbar\right)^3}\delta \left( \epsilon_p + \epsilon_{p_1} - \epsilon_{p_2} - \epsilon_{p_3} \right) \delta \left(\textbf{p} + \textbf{p}_1-\textbf{p}_2 - \textbf{p}_3\right) \nonumber \\
&& \ \ \ \ \ \ \ \ \ \ \ \ \ \ \ \ \ \ \ \ \times \left[ \left\{ \uuline{1} + \uuline{W} \left( \textbf{p}\right) ,\uuline{\Sigma}^< \left( \textbf{p}_1,\textbf{p}_2,\textbf{p}_3 \right) \right\} - \left\{ \uuline{W} \left( \textbf{p}\right) ,\uuline{\Sigma}^> \left( \textbf{p}_1,\textbf{p}_2,\textbf{p}_3 \right)\right\} \right] .\label{1001}
\end{eqnarray}

\section{Formulation of the moment equations}\label{Appendix_Moment}
In the linearized kinetic equation Eq.(\ref{10}), simplified notations for the matrices are defined by
\begin{eqnarray}
\begin{array}{l}
\uuline{\delta W}^A\equiv\begin{pmatrix}
	0 & -\delta W_{10} & -2\delta W_{1-1}\\
	\delta W_{01} & 0 & -\delta W_{0-1}\\
	2\delta W_{-11} & \delta W_{-10} & 0
\end{pmatrix}, \ \ \ \  
\uuline{\delta W}^B\equiv\begin{pmatrix}
	0 & -\delta W_{10} & 0\\
	\delta W_{01} & 0 & \delta W_{0-1}\\
	0 & -\delta W_{-10} & 0
\end{pmatrix},\\
\\
\uuline{\delta W}^C\equiv\begin{pmatrix}
	0 & -\delta W_{10} & -\delta W_{1-1}\\
	\delta W_{01} & 0 & 0\\
	\delta W_{-11} & 0 & 0
\end{pmatrix}, \ \ \ \  
\uuline{\delta W}^D\equiv\begin{pmatrix}
	0 & -\delta W_{10} & -3\delta W_{1-1}\\
	\delta W_{01} & 0 & -2\delta W_{0-1}\\
	3\delta W_{-11} & 2\delta W_{-10} & 0
\end{pmatrix},\\
\\
\uuline{\delta n}^C\equiv\begin{pmatrix}
	0 & -\delta n_{10} & \delta n_{1-1}\\
	-\delta n_{01} & 0 & 0\\
	-\delta n_{-11} & 0 & 0
\end{pmatrix}, \ \ \ \  
\uuline{\delta n}^D\equiv\begin{pmatrix}
	0 & \delta n_{10}+2\delta n_{0-1} & -\delta n_{1-1}\\
	-\delta n_{01}-2\delta n_{-10} & 0 & 0\\
	\delta n_{-11} & 0 & 0
\end{pmatrix}.
\end{array}
\end{eqnarray}

From Eq.(\ref{10}), one can derive a general moment equation. First, we consider the precession mode by inserting $\chi_0=1$ into the general moment equation. The moment equation for the precession mode is given by Eq.(\ref{A1}), where several matrices are defined as
\begin{eqnarray}
\begin{array}{l}
\uuline{\langle \chi_0\rangle}^A\equiv\begin{pmatrix}
	0 & \langle\chi_0\rangle_{10} & 2\langle\chi_0\rangle_{1-1}\\
	-\langle\chi_0\rangle_{01} & 0 &\langle\chi_0\rangle_{0-1}\\
	-2\langle\chi_0\rangle_{-11} & -\langle\chi_0\rangle_{-10} & 0
\end{pmatrix}, \ \ \ \   
\uuline{\langle \chi_0\rangle}^B\equiv\begin{pmatrix}
	0 & \langle\chi_0\rangle_{10} & 0\\
	-\langle\chi_0\rangle_{01} & 0 &-\langle\chi_0\rangle_{0-1}\\
	0 & \langle\chi_0\rangle_{-10} & 0
\end{pmatrix},\\
\\
\uuline{\langle \chi_0\rangle}^D\equiv\begin{pmatrix}
	0 & -\langle\chi_0\rangle_{0-1} & 2\langle\chi_0\rangle_{1-1}\\
	\langle\chi_0\rangle_{-10} & 0 &\langle\chi_0\rangle_{0-1}\\
	-2\langle\chi_0\rangle_{-11} & -\langle\chi_0\rangle_{-10} & 0
\end{pmatrix}.
\end{array}
\end{eqnarray} 

We next consider the dipole mode by inserting $\chi_1=x$ and $\chi_2=p_x/m$ into the general moment equation. The moment equations are given in Eqs.(\ref{14}) and (\ref{15}), where several matrices are defined as 
\begin{eqnarray}
\begin{array}{l}
\uuline{\langle \chi_1\rangle}^A\equiv\begin{pmatrix}
	0 & \langle\chi_1\rangle_{10} & 2\langle\chi_1\rangle_{1-1}\\
	-\langle\chi_1\rangle_{01} & 0 &\langle\chi_1\rangle_{0-1}\\
	-2\langle\chi_1\rangle_{-11} & -\langle\chi_1\rangle_{-10} & 0
\end{pmatrix}, \ \ \ \   
\uuline{\langle \chi_1\rangle}^B\equiv\begin{pmatrix}
	0 & \langle\chi_1\rangle_{10} & 0\\
	-\langle\chi_1\rangle_{01} & 0 &-\langle\chi_1\rangle_{0-1}\\
	0 & \langle\chi_1\rangle_{-10} & 0
\end{pmatrix},\\
\\
\uuline{\langle \chi_1\rangle}^D\equiv\begin{pmatrix}
	0 & -\langle\chi_1\rangle_{0-1} & 2\langle\chi_1\rangle_{1-1}\\
	\langle\chi_1\rangle_{-10} & 0 &\langle\chi_1\rangle_{0-1}\\
	-2\langle\chi_1\rangle_{-11} & -\langle\chi_1\rangle_{-10} & 0
\end{pmatrix},
\end{array}
\end{eqnarray} 
and 
\begin{eqnarray}
\begin{array}{l}
\uuline{\langle \chi_2\rangle}^A\equiv\begin{pmatrix}
	0 & \langle\chi_2\rangle_{10} & 2\langle\chi_2\rangle_{1-1}\\
	-\langle\chi_2\rangle_{01} & 0 &\langle\chi_2\rangle_{0-1}\\
	-2\langle\chi_2\rangle_{-11} & -\langle\chi_2\rangle_{-10} & 0
\end{pmatrix}, \ \ \ \   
\uuline{\langle \chi_2\rangle}^B\equiv\begin{pmatrix}
	0 & \langle\chi_2\rangle_{10} & 0\\
	-\langle\chi_2\rangle_{01} & 0 &-\langle\chi_2\rangle_{0-1}\\
	0 & \langle\chi_2\rangle_{-10} & 0
\end{pmatrix},\\
\\
\uuline{\langle \chi_2\rangle}^C\equiv\begin{pmatrix}
	0 & \langle\chi_2\rangle_{10} & \langle\chi_2\rangle_{1-1}\\
	-\langle\chi_2\rangle_{01} & 0 &0\\
	-\langle\chi_2\rangle_{-11} & 0 & 0
\end{pmatrix},\ \ \ \ 
\uuline{\langle \chi_2\rangle}^D\equiv\begin{pmatrix}
	0 & \langle\chi_2\rangle_{10} & 3\langle\chi_2\rangle_{1-1}\\
	-\langle\chi_2\rangle_{01} & 0 &2\langle\chi_2\rangle_{0-1}\\
	-3\langle\chi_2\rangle_{-11} & -2\langle\chi_1\rangle_{-10} & 0
\end{pmatrix}.
\end{array}
\end{eqnarray}

\section{Relaxation rates}\label{Appendix_Relaxation}
In this appendix we consider the linearized collision integral $\left.\partial \delta\uuline{W}/{\partial t}\right|_{coll}$ appearing in the linearized kinetic equation Eq.(\ref{10}). Assuming small amplitude spin oscillations around fully polarized state, we divide $\uuline{W}$ into the initial part and the fluctuation part:
\begin{eqnarray}
\uuline{W}\left(\textbf{p},\textbf{r},t\right)=\uuline{W}^0\left(\textbf{p},\textbf{r}\right)+\delta\uuline{W}\left(\textbf{p},\textbf{r},t\right),\label{Appendix_A_1}
\end{eqnarray}   
where the initial distribution $\uuline{W}^0$ is given by Eqs.(\ref{equilibrium_distribution}) and (\ref{for_Appendix B}). For the linearized form of the collision term, it is convenient to express the fluctuation of the distribution function as
\begin{eqnarray}
\delta\uuline{W}\left(\textbf{p},\textbf{r},t\right)=f_0\left(\textbf{p},\textbf{r}\right)\uuline{\phi}\left(\textbf{p},\textbf{r},t\right),\label{Appendix_A_2}
\end{eqnarray}
where $f_0\left(\textbf{p},\textbf{r}\right)$ is given in Eq.(\ref{for_Appendix B}). The explicit expression for $ij$ component of Eq.(\ref{Appendix_A_2}) can be related back to the set of moments using Eq.(\ref{chi}) as
\begin{eqnarray}
\phi_{ij} = \langle \chi_0 \rangle _{ij} +  \beta m\omega_x^2 \langle \chi_1 \rangle_{ij}x + \beta \langle \chi_2 \rangle_{ij} p_x.
\end{eqnarray}
 
Inserting Eq.(\ref{Appendix_A_1}) into Eq.(\ref{1001}) and linearizing it in $\uuline{\phi}$, we obtain the linearized collision integral $\left.\partial \delta\uuline{W}/{\partial t}\right|_{coll}$:

\begin{eqnarray}
\frac{\partial \delta W_{ij}}{\partial t}\mid_{coll}&=& \frac{\pi}{\hbar }  \frac{1}{\left( 2\pi \hbar \right)^6}\sum_{ll^\prime m i^\prime i^ {\prime\prime}j^\prime j^{\prime\prime}}\int d\textbf{p}_1\int d\textbf{p} _2 \int d\textbf{p} _3 \delta \left( \epsilon_p + \epsilon_{p_1} - \epsilon_{p_2} - \epsilon_{p_3} \right) \delta {\left( \textbf{p} + \textbf{p}_1-\textbf{p}_2 - \textbf{p}_3 \right) } \nonumber \\
 & & \ \ \ \ \ \ \ \ \ \ \ \ \ \ \ \ \times\Biggl\{4f_0 \left( \textbf{p}_2 \right) f_0 \left( \textbf{p}_3 \right)V_{11mj}V_{mi11} \Biggr. \nonumber \\
 & & \ \ \ \ \ \ \ \ \ \ \ \ \ \ \ \ \ \ \ -2f_0 \left( \textbf{p} \right) f_0 \left( \textbf{p}_1 \right)\Bigl[V_{ll^\prime 1j} V_{11l^\prime l} \delta_{1i} +V_{i1ll^\prime} V_{l^\prime l11} \delta_{1j}\Bigr] \nonumber \\ 
 & & \ \ \ \ \ \ \ \ \ \ \ \ \ \ \ \ \ \ \ \ + f_0 \left( \textbf{p}_2 \right) f_0 \left( \textbf{p}_3 \right) \Bigl[V_{1l^\prime mj}V_{mij^\prime 1} \phi _{j^\prime l^\prime} \left( \textbf{p}_2 \right)+ V_{1l^\prime mj}V_{mi1j^{\prime\prime}} \phi _{j^{\prime\prime} l^\prime} \left( \textbf{p}_2 \right) \Bigr. \nonumber \\
& & \ \ \ \ \ \ \ \ \ \ \ \ \ \ \ \ \ \ \ \ \ \ \ \ \ \ \ \ \ \ \ \ \ \ \ \ \ \ \ \ + V_{im1l^\prime} V_{i^\prime 1mj} \phi _{l^\prime i^\prime} \left( \textbf{p}_2 \right) + V_{im1l^\prime} V_{1i^{\prime\prime} mj} \phi _{l^\prime i^{\prime\prime}} \left( \textbf{p}_2 \right)  \nonumber \\
& & \ \ \ \ \ \ \ \ \ \ \ \ \ \ \ \ \ \ \ \ \ \ \ \ \ \ \ \ \ \ \ \ \ \ \ \ \ \ \ \ + V_{l1mj}V_{mi1j^{\prime\prime}} \phi _{j^{\prime\prime} l} \left( \textbf{p}_3 \right) + V_{l1mj}V_{mij^\prime 1} \phi _{j^{\prime} l} \left( \textbf{p}_3 \right) \nonumber \\
& & \ \ \ \ \ \ \ \ \ \ \ \ \ \ \ \ \ \ \ \ \ \ \ \ \ \ \ \ \ \ \ \ \ \ \ \ \ \ \ \ \Bigl. + V_{iml1} V_{1i^{\prime\prime}mj} \phi _{l i^{\prime\prime}} \left( \textbf{p}_3 \right) + V_{iml1} V_{i^\prime mj} \phi _{li^\prime} \left( \textbf{p}_3 \right) \Bigr] \nonumber \\
 & & \ \ \ \ \ \ \ \ \ \ \ \ \ \ \ \ \ \ \ \ + f_0 \left( \textbf{p} \right) f_0 \left( \textbf{p}_1 \right) \Bigl[ V_{ll^\prime 1j} \left( V_{1 i^{\prime\prime}l^\prime l} + V_{1i^{\prime\prime}ll^\prime} \right) \phi _{i i^{\prime\prime}} \left( \textbf{p} \right) \Bigr. \nonumber \\
& & \ \ \ \ \ \ \ \ \ \ \ \ \ \ \ \ \ \ \ \ \ \ \ \ \ \ \ \ \ \ \ \ \ \ \ \ \ \ \ + V_{i1ll^\prime} \left( V_{l^\prime l1 j^{\prime\prime}} + V_{ll^\prime 1j^{\prime\prime}} \right) \phi _{j^{\prime\prime}j} \left( \textbf{p} \right) \nonumber \\
& & \ \ \ \ \ \ \ \ \ \ \ \ \ \ \ \ \ \ \ \ \ \ \ \ \ \ \ \ \ \ \ \ \ \ \ \ \ \ \ + V_{ll^\prime mj} \left( V_{i^\prime 1l^\prime l} + V_{i^\prime 1ll^\prime} \right) \delta_{1i} \phi _{mi^{\prime}} \left( \textbf{p}_1 \right) \nonumber \\
& & \ \ \ \ \ \ \ \ \ \ \ \ \ \ \ \ \ \ \ \ \ \ \ \ \ \ \ \ \ \ \ \ \ \ \ \ \ \ \ \Biggl.\Bigl. + V_{imll^\prime} \left( V_{l^\prime l j^{\prime}1} + V_{ll^\prime 1j^{\prime}1} \right) \delta_{1j} \phi _{j^{\prime}m} \left( \textbf{p}_1 \right) \Bigr] \Biggr\}.\label{Appendix_B_4}
\end{eqnarray}

Using Eq.(\ref{13}), one can now express the moment associated with collisions $\langle\chi_\alpha\rangle_{ij}^{coll}$ in terms of the various relaxation rates:
\begin{eqnarray}
\langle\chi_\alpha\rangle_{ij}^{coll}\equiv \sum_{\beta}\sum_{lm}\left(c^{\alpha\beta}_0\gamma_0+c^{\alpha\beta}_1\gamma_1+c^{\alpha\beta}_2\gamma_2\right)\langle\chi_\beta\rangle_{lm}.
\end{eqnarray}
In above equation, the sum $\sum_\beta$ implies that the moment equations couple lower moments to higher moments through $\langle\chi_\alpha\rangle_{ij}^{coll}$. In addition, the sum $\sum_{lm}$ implies that the collision term mixes different spin states. The coefficients $c_i^{\alpha\beta}$ should be determined by taking moments of the linearized collision integral Eq.(\ref{Appendix_B_4}), and the  relaxation rates $\gamma_0$, $\gamma_1$, and $\gamma_2$ are defined as
\begin{eqnarray}
\gamma_0 &\equiv&  \frac{\pi}{\hbar }  \frac{1}{\left( 2\pi \hbar \right)^9}\frac{g_2^2}{N} \int d\textbf{r} \int d\textbf{p} \int d\textbf{p}_1\int d\textbf{p} _2 \int d\textbf{p} _3 \nonumber \\
	& & \ \ \ \ \ \ \ \ \ \ \ \ \ \ \ \ \times \delta \left( \epsilon_p + \epsilon_{p_1} - \epsilon_{p_2} - \epsilon_{p_3} \right) \delta {\left( \textbf{p} + \textbf{p}_1-\textbf{p}_2 - \textbf{p}_3 \right) } f_0 \left( \textbf{p} \right)f_0 \left( \textbf{p}_1 \right),\label{gamma00}
\end{eqnarray}
\begin{eqnarray}
\gamma_1 &\equiv&  \frac{\pi}{\hbar }  \frac{1}{\left( 2\pi \hbar \right)^9}\frac{g_1g_2}{N} \int d\textbf{r} \int d\textbf{p} \int d\textbf{p}_1\int d\textbf{p} _2 \int d\textbf{p} _3 \nonumber \\
	& & \ \ \ \ \ \ \ \ \ \ \ \ \ \ \ \ \times \delta \left( \epsilon_p + \epsilon_{p_1} - \epsilon_{p_2} - \epsilon_{p_3} \right) \delta {\left( \textbf{p} + \textbf{p}_1-\textbf{p}_2 - \textbf{p}_3 \right) } f_0 \left( \textbf{p} \right)f_0 \left( \textbf{p}_1 \right),\label{gamma11}
\end{eqnarray}
\begin{eqnarray}
\gamma_2 &\equiv&  \frac{\pi}{\hbar }  \frac{1}{\left( 2\pi \hbar \right)^9}\frac{g_0^2}{N} \int d\textbf{r} \int d\textbf{p} \int d\textbf{p}_1\int d\textbf{p} _2 \int d\textbf{p} _3 \nonumber \\
	& & \ \ \ \ \ \ \ \ \ \ \ \ \ \ \ \ \times \delta \left( \epsilon_p + \epsilon_{p_1} - \epsilon_{p_2} - \epsilon_{p_3} \right) \delta {\left( \textbf{p} + \textbf{p}_1-\textbf{p}_2 - \textbf{p}_3 \right) } f_0 \left( \textbf{p} \right)f_0 \left( \textbf{p}_1 \right).\label{gamma22}
\end{eqnarray}
We estimate these relaxation rates, assuming the number of atoms $N=1\times10^5$, trap frequencies $\left\{\omega_x/2\pi,\omega_y/2\pi,\omega_z/2\pi\right\}=\left\{24,240,240\right\} {\rm Hz}$ and the magnetic field $B=20\times10^{-7} {\rm T}$. We find $\gamma_0=3.0\times10^{-4} s^{-1}$, $\gamma_1=-6.5\times10^{-2} s^{-1}$, and $\gamma_2=14 s^{-1}$ for ${\rm ^{87}Rb}$, while $\gamma_0=1.0\times10^{-2} s^{-1}$, $\gamma_1=3.1\times10^{-1} s^{-1}$, and $\gamma_2=9.7 s^{-1}$ for ${\rm ^{23}Na}$, respectively.


\end{document}